\documentclass[modern]{aastex62}
\usepackage{color}

\newcommand{\is}{i_{\star}}
\newcommand{\io}{i_{\rm orb}}
\newcommand{\spl}{\delta\nu_{\star}}
\newcommand{\vsini}{v_{\rm rot} \sin{i_{\star}}}

\graphicspath{{./}{figures/}}

\received{2018 November 23}
\revised{2019 February 1}
\accepted{2019 February 5}
\submitjournal{AJ}

\shorttitle{Misaligned orbit of Kepler-408b}
\shortauthors{Kamiaka et al.}

\begin{document}

\title{The misaligned orbit of the Earth-sized planet Kepler-408b}

\correspondingauthor{Yasushi Suto}
\email{suto@phys.s.u-tokyo.ac.jp}

\author[0000-0001-6036-3194]{Shoya Kamiaka}
\affiliation{Department of Physics, The University of Tokyo, Tokyo,
  113-0033, Japan}

\author[0000-0001-9405-5552]{Othman Benomar}
\affiliation{Center for Space Science, NYUAD Institute, New York
  University Abu Dhabi, PO Box 129188, Abu Dhabi, UAE}

\author[0000-0002-4858-7598]{Yasushi Suto}
\affiliation{Department of Physics, The University of Tokyo, Tokyo,
  113-0033, Japan}
\affiliation{Research Center for the Early Universe, School of
  Science, The University of Tokyo, Tokyo 113-0033, Japan}

\author[0000-0002-8958-0683]{Fei Dai}
\affiliation{Department of Physics and Kavli Institute for
  Astrophysics and Space Research, Massachusetts Institute of
  Technology, Cambridge, MA 02139, USA}
\affiliation{Department of Astrophysical Sciences, Princeton
  University, Princeton, NJ 08544, USA}

\author[0000-0003-1298-9699]{Kento Masuda}
\affiliation{Department of Astrophysical Sciences, Princeton
  University, Princeton, NJ 08544, USA}
\affiliation{NASA Sagan Fellow}

\author[0000-0002-4265-047X]{Joshua N.\ Winn}
\affiliation{Department of Astrophysical Sciences, Princeton
  University, Princeton, NJ 08544, USA}

\begin{abstract}
  Kepler-408 is one of the 33 planet-hosting {\it Kepler} stars
  for which asteroseismology has been used to investigate
  the orientation of the stellar rotation axis relative to the
  planetary orbital plane.  The transiting ``hot Earth,'' Kepler-408b,
  has an orbital period of 2.5 days and a radius of $0.86$~$R_\oplus$,
  making it much smaller than the planets for which spin-orbit
  alignment has been studied using the Rossiter-McLaughlin effect.
  Because conflicting asteroseismic results have been reported in the
  literature, we undertake a thorough re-appraisal of this system
  and perform numerous checks for consistency and robustness. We
  find that the conflicting results are due to the different models for the 
  low-frequency noise in the power spectrum.  A careful
  treatment of the background noise resolves these conflicts,
  and shows that the stellar inclination is $\is=42^{+5}_{-4}$ degrees.
  Kepler-408b is, by far, the smallest planet known to
  have a significantly misaligned orbit.
\end{abstract}

\keywords{asteroseismology --- stars: oscillations --- stars: rotation --- stars: planetary systems --- methods: data analysis --- techniques: photometric}

\vspace*{1cm}

\section{Introduction} \label{sec:intro}

Planets around other stars are occasionally found to have orbits
that are misaligned, or even retrograde, relative to the direction of
stellar rotation \citep[e.g.,][]{WinnFabrycky2015,Triaud2017}.
However, all previous detections of misaligned orbits are for planets
larger than Neptune.  Smaller planets are relatively unexplored
because of the difficulty of the relevant measurements.

Three of the techniques for investigating spin-orbit alignment --- the
Rossiter-McLaughlin effect, the starspot-tracking method, and the
gravity-darkening method --- require the observation of signals for
which the amplitude is proportional to the loss of light during
planetary transits.  Hence, they are much easier to apply to giant
planets than small planets.  Two other techniques --- the
asteroseismic method, and the $v\sin i$ method --- rely on observing
signals that are independent of planet size.  However, the
asteroseismic method has only been applied to 33 stars, because it
requires an unusually bright star with large-amplitude $p$-mode
oscillations.  The $v\sin i$ method has been applied to samples of
hundreds of stars, but in most cases it only provides weak constraints
\citep{Schlaufman2010,Winn+2017}.  Due to these limitations, it is
unclear whether the misalignments are the result of processes specific
to giant planets, or whether they also occur for terrestrial planets.

Kepler-408 (also known as KIC\,10963065 and KOI-1612) is one of
approximately 150{,}000 Sun-like stars that were monitored for 4 years
with the NASA {\it Kepler} space telescope \citep{Borucki+2010}.  Its
lightcurves exhibit a periodic transit signal due to an Earth-sized
planet with $P_{\rm orb}\sim2.5$ days \citep{Marcy2014}.  Table
\ref{tab:parameters} summarizes the known characteristics of the
system.  With a {\it Kepler} apparent magnitude of 8.8, the host star
is the third brightest of all the {\it Kepler} stars with confirmed
planets.  This unusual brightness enables an investigation of the
stellar obliquity using asteroseismology.  In particular, it is
possible to determine the inclination $\is$ of the stellar rotation
axis based on the fine structure in the $p$-mode pulsation spectrum
\citep{Toutain1993,Gizon2003}.

\begin{deluxetable}{lcl}
\label{tab:parameters}
\tablewidth{0pt}
\tablecolumns{2}
\tablecaption{System Parameters of Kepler-408\label{tab:params}}
\tablehead{\colhead{Parameter} & \colhead{Value} & \colhead{Reference}
	}
\startdata
{\it Stellar Parameters} \\ \hline
Effective temperature, $T_{\rm eff}$~[K] & $6088\pm 65$ &
\citet{Petigura+2017} \\
Surface gravity, $\log (g/{\rm cm\,s}^{-2})$        &  $4.318^{+0.08}_{-0.089}$    &  \citet{Petigura+2017} \\
Metallicity, [Fe/H]                                  &  $-0.138^{+0.043}_{-0.042}$  &  \citet{Petigura+2017} \\
Mass $M_\star$ [$M_\odot$] & $1.05\pm 0.04$ & \citet{Johnson+2017} \\
Radius $R_\star$ [$R_\odot$] & $1.253\pm 0.051$ & \citet{Berger+2018}
\\
Age [Gyr] & $4.7 \pm 1.2$ & \citet{Johnson+2017} \\
Projected rotation rate, $\vsini$ [km~s$^{-1}$] & $2.8\pm 1.0$ &
\citet{Petigura+2017} \\
Rotation period $P_{\rm rot}$ [days] & $12.89\pm 0.19$ &
\citet{Angus+2018} \\[0.1in]
\hline
{\it Planetary Parameters} \\ \hline
Planet-to-star radius ratio, $R_{\rm p}/R_\star$ & $0.0063\pm
        0.0003$ & This work \\
Radius $R_{\rm p}$ [$R_\oplus$] & $0.86\pm 0.04$ & This work\\
Time of inferior conjunction [BJD] & $2454965.6804 \pm 0.0003$ & This
work \\
Orbital period $P_{\rm orb}$ [days] & $2.465024 \pm 0.000005$ &
\citet{Thompson+2018} \\
Orbital inclination $\io$ [deg] & $81.85 \pm 0.10$ & This work \enddata
\end{deluxetable}

However, there are conflicting reports in the literature.
\citet{Campante+2016} found the inclination to be consistent with
$90^\circ$ and set a lower limit of $54^{\circ}$.  This was part of a
homogeneous study of 25 stars with transiting planets.  In contrast,
\citet{Nielsen+2017} found the inclination to be between 40 and 45
degrees.  This finding was incidental to the main purpose of the
study, which was to probe the internal rotation profiles of 6 stars.
The authors did not remark on the transiting planet, nor on the
conflict with \citet{Campante+2016}.

We have examined the case of Kepler-408 in greater detail, to try and
resolve this conflict.  We were also motivated by the numerical
simulations of \citet{Kamiaka+2018}, who established the observational
requirements for the reliable inference of the rotational inclination,
and found that the characteristics of Kepler-408 should allow for
reliable results.  Section \ref{sec:transit} describes the transit
analysis.  Section \ref{sec:rotation} presents some independent checks
on the previous measurements of the stellar rotation period, which
plays a key role in the asteroseismic analysis.  Section
\ref{sec:seismology} describes the asteroseismic analysis, and
resolves the prior discrepancy by identifying a problem with the
analysis by \citet{Campante+2016}. Section \ref{sec:spectroscopy}
shows that our asteroseismic estimate of $\is$ agrees with the
constraint that is obtained by combining measurements of the stellar
radius, rotation period, and sky-projected rotation velocity.  Our
findings and some implications are summarized in Section
\ref{sec:summary}. Just for definiteness, the present paper refers to
those systems as {\it misaligned} if either $\lambda$ (sky-projected spin-orbit angle) or $90^\circ -\is$ (a proxy for the stellar obliquity in transiting planetary systems)
exceeds $30^{\circ}$ in 95\% confidence.

\section{Transit modeling} \label{sec:transit}

\begin{figure}[ht]
\centering
\includegraphics[width=0.8\columnwidth]{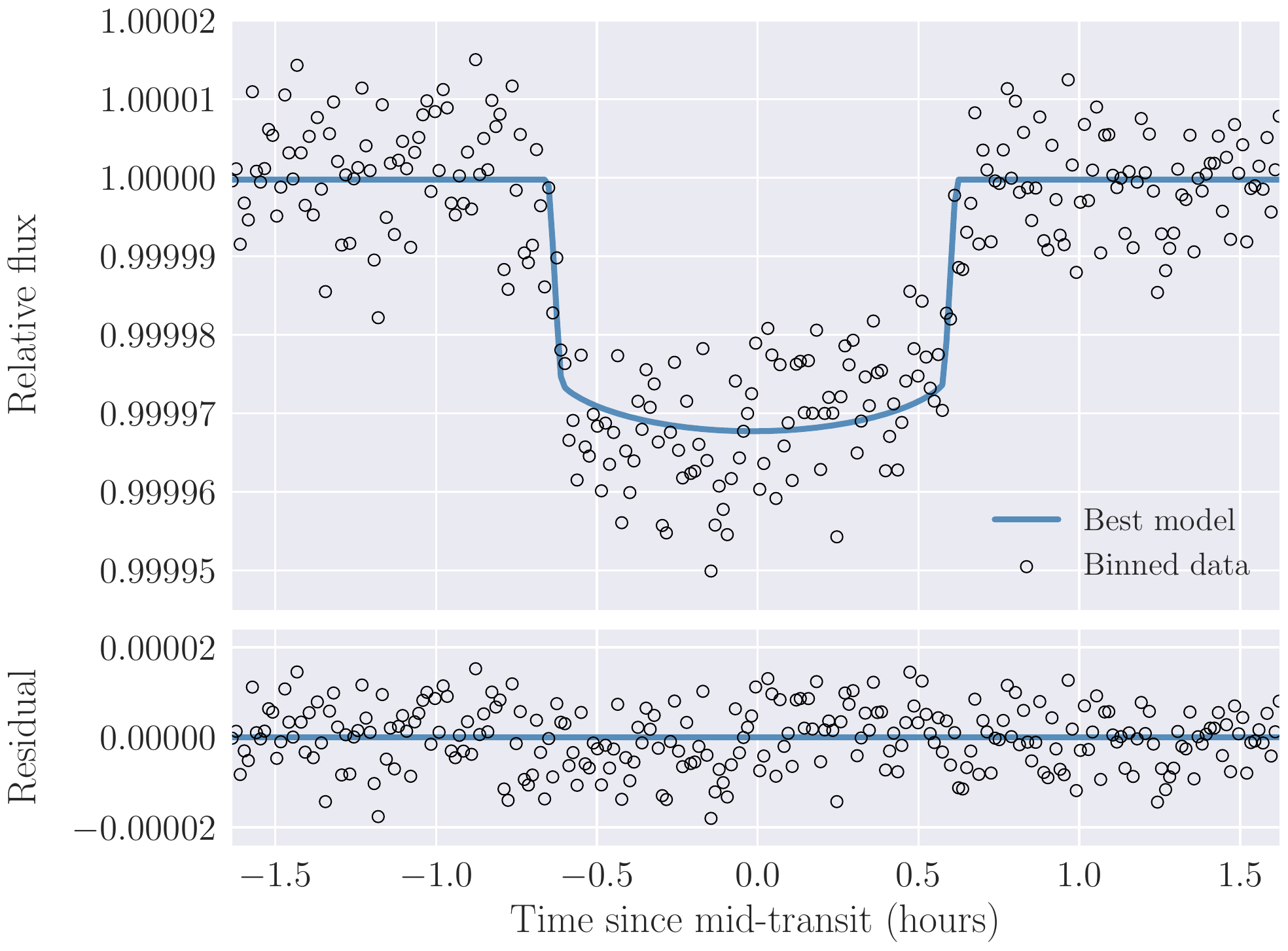}
\caption{ Phase-folded transit light curve of Kepler-408b.  {\it Upper
    panel} --- Binned data (open circles) with the best-fitting model
  (thick line). {\it Lower panel} --- Residuals between the data and
  the best-fitting model.}
\label{fig:transit-kepler408b}
\end{figure}

The orbital inclination, $\io$, of a transiting planet is always close
to 90$^\circ$. For a precise measurement, we modeled the {\it Kepler}
transit light curve. We downloaded the short-cadence, pre-search data
conditioning (PDC) light curves from the Mikulski Archive for Space
Telescopes. The data surrounding each transit were fitted with a
standard model for the loss of light \citep{MandelAgol2002}, assuming
the orbit to be circular and accounting for stellar variability with a
locally quadratic function of time.  After dividing through by the
best-fitting quadratic functions, the transit data were phase-folded
and averaged, giving a mean light curve with a higher signal-to-noise
ratio (Figure~\ref{fig:transit-kepler408b}).

This light curve was then fitted to obtain our final estimates for the
transit parameters (Table~\ref{tab:parameters}).  Uniform priors were
adopted for the logarithm of the planet-to-star radius ratio ($R_{\rm
  p}/R_\star$), the cosine of the orbital inclination ($\cos \io$),
the normalization of the light curve, the two coefficients of the
quadratic limb-darkening profile, and the logarithm of a noise term to
account for the scatter of the residuals between the data and the
model. A Gaussian prior was adopted for the mean stellar density
($\rho_\star=0.816\pm0.025~\mathrm{g~cm^{-3}}$) based on the previous
asteroseismic analysis of \citet{Kamiaka+2018}. The posterior
distributions for the model parameters were obtained with a nested
sampling code \citep{Feroz+2009}. The result for the orbital
inclination was $\io = 81.85 \pm 0.10$~degrees.

In the present analysis, we adopted the circular model because the
time-scale for tidal orbital circularization is likely short for the
2.5-day orbit. We also checked that the model with non-zero
eccentricity does not significantly improve the fit. This is in
agreement with the analysis of \citet{VanEylen+2018}, who found that
the eccentricity was consistent with zero within $95\%$ confidence.

\section{Stellar rotation period from photometric variability}
\label{sec:rotation}

The {\it Kepler} photometric time series exhibits quasi-periodic
modulation that is presumably due to the rotation of surface
inhomogeneities across the star's visible hemisphere.  By computing
the autocorrelation function, \citet{McQuillan+2013} determined the
photometric rotation period to be $12.44\pm 0.17$~days.
\citet{Angus+2018} reported a value of $12.89\pm 0.19$~days by
modeling the {\it Kepler} data as a Gaussian process with a
quasi-periodic covariance kernel function.

\begin{figure}[ht!]
\centering
\centerline{\includegraphics[width=\columnwidth]{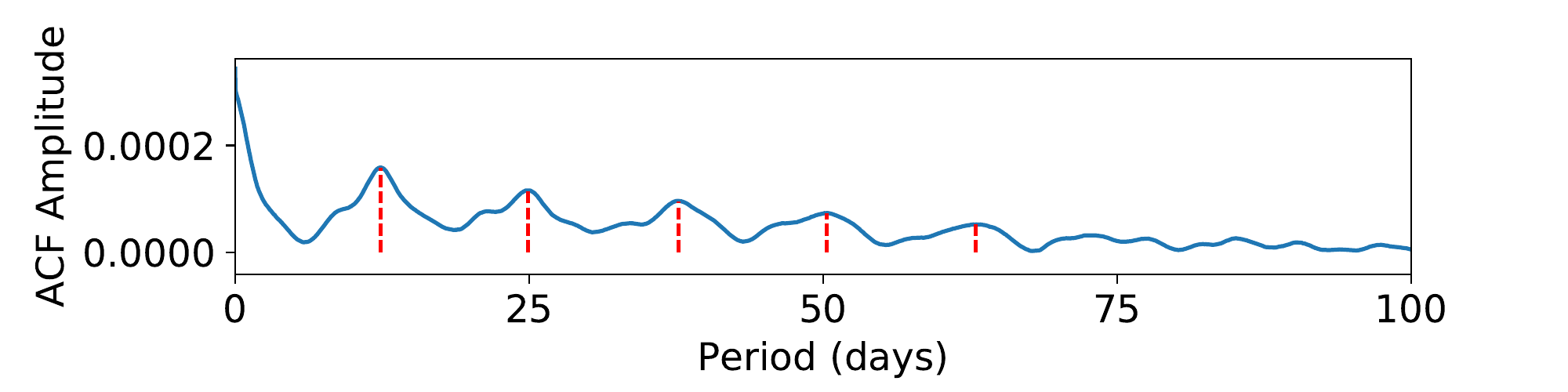}}
\centerline{\includegraphics[width=\columnwidth]{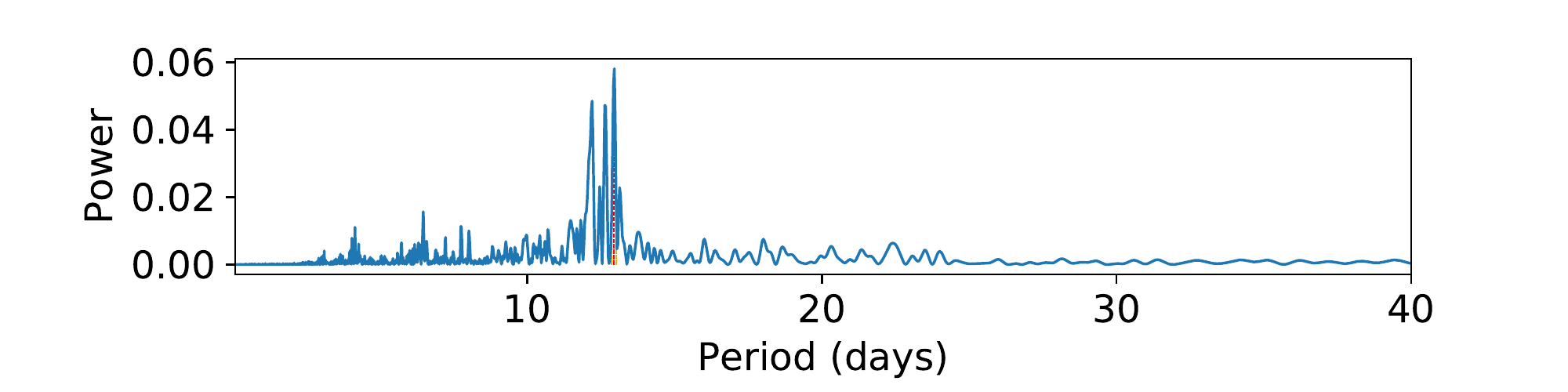}}
\centerline{\includegraphics[width=\columnwidth]{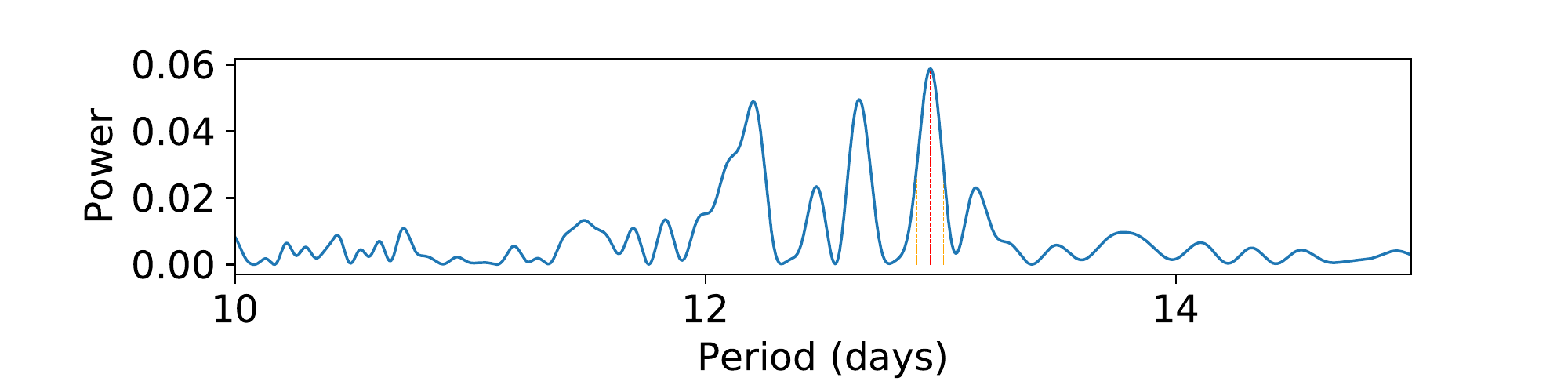}}
\caption{Estimates of the stellar rotation period from the
    photometric time series.  {\it Top:} Auto-correlation function. The
    red dashed lines indicate the locations of several peaks.  {\it
      Middle:} The Lomb-Scargle periodogram. The location of the most
    significant peak is marked with a red dashed line, and the
    $1\sigma$ uncertainty interval is plotted with orange dashed lines.  {\it
      Bottom:} Close-up of the Lomb-Scargle periodogram around the
    most significant peak.}
\label{fig:acf-periodogram}
\end{figure}

To perform an independent check on the determination of the stellar
rotation period, we analyzed the {\it Kepler} data outside of
transits. We normalized the data from each quarter by setting the
median flux equal to unity. A Lomb-Scargle periodogram of the
resulting time series has its most prominent peak at $12.96\pm
0.07$~days, and the autocorrelation function shows a series of peaks
spaced by $12.94 \pm 0.22$ days (Figure~\ref{fig:acf-periodogram}).
Previous experience has shown that the strongest photometric
periodicity sometimes occurs at harmonics of the true rotation period,
presumably because there are several active regions on the star.

\begin{figure}[hb!]
\centering
\includegraphics[width=0.8\columnwidth]{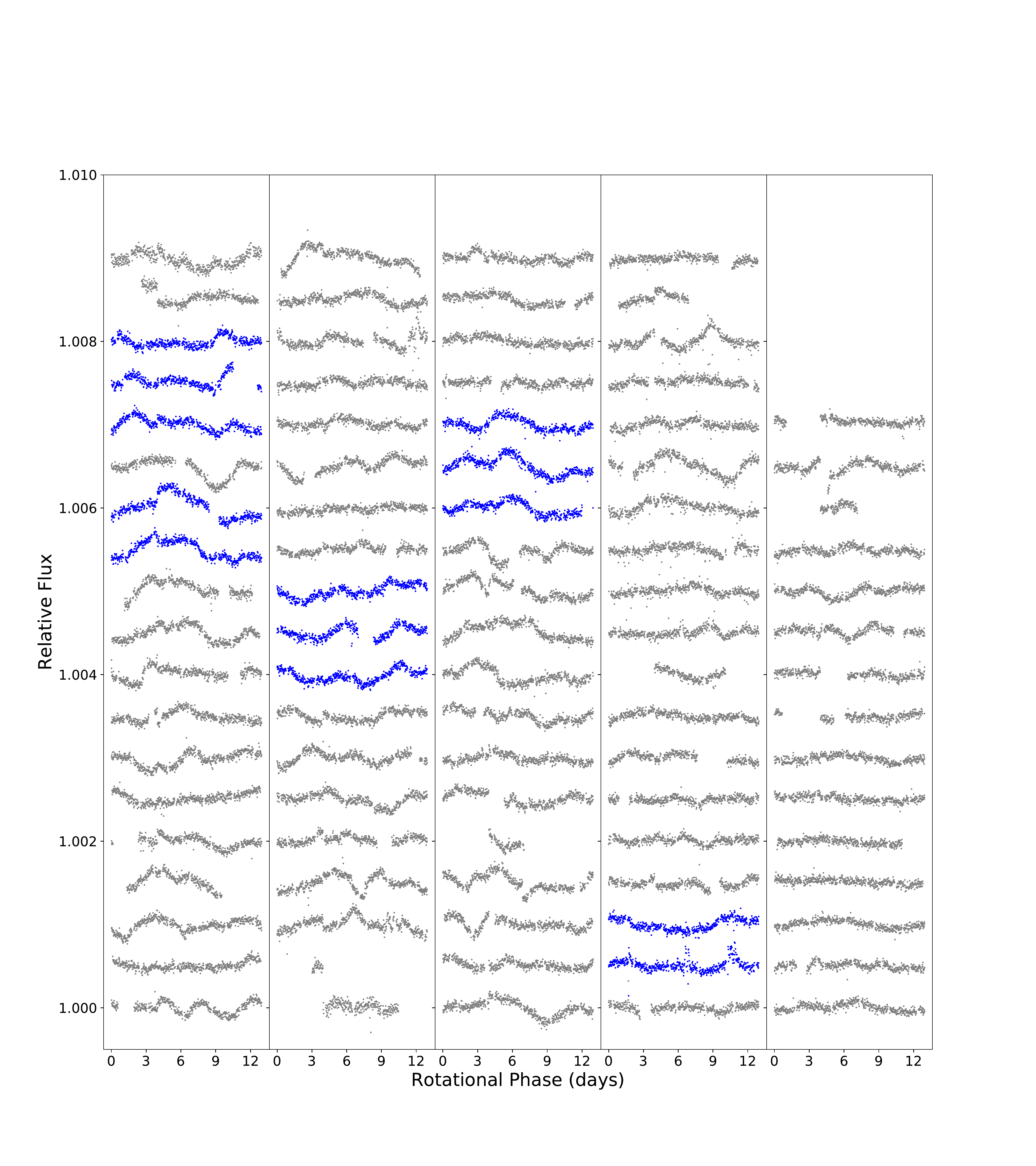}
\caption{Confirmation of the rotation period through visual inspection.
  Shown is the entire {\it Kepler} light curve, folded with the candidate
  12.94-day period. Vertical offsets have been applied to each cycle of data
  in order to separate them; they are organized like 
  the lines of text on a page.
  Highlighted in blue are several occasions where the pattern
  of flux variation is similar from one rotation to the
  next.  This would be unlikely if the rotation period had been
  misidentified.}
\label{fig:fold}
\end{figure}

In the present case, visual inspection of the light curve confirms
that the true period is close to 12.9 days.  We were able to identify
several time intervals in which a complex pattern of variations
repeats nearly exactly after 12.9 days (Figure~\ref{fig:fold}), which
would be an unlikely coincidence if the true period were different.
We highlighted part of the lightcurves in Figure~\ref{fig:fold} so as
to clarify the periodicity, but it should be regarded as a {\it sanity
  check} on the more objective measures with no claim to be objective
or complete.  For a more systematic comparison between asteroseismic
  and photometric estimates of stellar rotation periods for other
  stars, see \citet{Suto2019}.

In what follows, we adopt the value $P_{\rm rot} = 12.89\pm 0.19$~days
based on the work of \citet{Angus+2018}, since their analysis appears
to be the most rigorous with regard to the quoted uncertainty.  The
reciprocal of the rotation period, which is most relevant to the
asteroseismic analysis, is $1/P_{\rm rot} = 0.898\pm 0.013~\mu$Hz.

\section{Asteroseismic analysis} \label{sec:seismology}

\subsection{Brief description of the method}

The star's pressure-mode oscillations ($p$ modes) are manifest in the
{\it Kepler} data as quasi-periodic variations in stellar brightness
with amplitudes of a few parts per million (ppm) and frequencies on
the order of 2000\,$\mu$Hz (periods~$\sim$~10\,min).  The modes can be
classified with three integers: the radial order $n (\geq 1)$, which
depends on the radial dependence of the oscillatory pattern; the
angular degree $\ell (\geq 0)$, which specifies the variation with the
polar angle; and the azimuthal order $m ( = -\ell...0...\ell)$, which
specifies the variation with the azimuthal angle.  For a non-rotating
star, all the modes with the same $n$ and $\ell$ would have the same
frequency, regardless of $m$. Rotation breaks this degeneracy,
producing small frequency shifts:
\begin{eqnarray}
  \label{eq:freq-m}
\nu_{n,\ell,m} = \nu_{n,\ell} + m\spl
\approx \left(n+\frac{\ell}{2}+ \varepsilon_{n,l}\right)\Delta\nu + m\spl,
\end{eqnarray}
where $\Delta\nu$ is the ``large separation'' (the spacing between
consecutive radial modes), $\varepsilon_{n,l}$ is a small correction
of order unity \citep{Tassoul1980, Tassoul1990, Mosser2013}, and
the rotational splitting $\spl$ is approximately the reciprocal
of the stellar rotation period \citep{Appourchaux2008}.

Because the shifts are small ($\spl \sim1\,\mu$Hz), all the modes
within a multiplet are expected to have the same intrinsic amplitude.
However, the height of each peak in the observed power spectrum is
also proportional to a factor $\mathcal{E}_{\ell,m}$ depending on the
rotational inclination $\is$ as shown in equations
(\ref{eq:profile-nlm}) and (\ref{eq:visibility}) below.  This is
because the peak heights are based on the average intensity of the
mode pattern across the visible hemisphere, and the modes have
different symmetries with respect to the rotation axis.

\subsection{Power spectrum modeling}

We downloaded the Kepler-408 power spectrum from the Kepler
Asteroseismic Science Operations Center database.  We modeled the
power spectrum as
\begin{eqnarray}
\label{eq:profile-nlm}
P(\nu) = 
\sum_{n=n_{\rm min}}^{n_{\rm max}}
\sum_{\ell=0}^{\ell_{\rm max}}
\sum_{m=-\ell}^{+\ell}~
\frac{H_{n,\ell}\,\mathcal{E}_{\ell,m}(\is)}
     {\displaystyle
       1 + \left( \frac{\nu-\nu_{n,\ell,m}} { \Gamma_{n,\ell,m}/2 } \right)^2 }
     + N(\nu),
\end{eqnarray}
where $H_{n,\ell}$ is the intrinsic mode amplitude,
$\mathcal{E}_{\ell,m}(\is)$ is the mode visibility (see
equation~\ref{eq:visibility}), $\nu_{n,\ell,m}$ is the line center,
$\Gamma_{n,\ell,m}$ is the line width, and $N(\nu)$ is the noise
background.  The background was modeled as
\begin{eqnarray}
\label{eq:noise-model}
N(\nu) = \frac{A_{1}}{1+(\tau_{1}\nu)^{p_{1}}} 
+ \frac{A_{2}}{1+(\tau_{2}\nu)^{p_{2}}} + N_{0},
\end{eqnarray}
where $N_0$ is a constant (white noise), and $A_i$, $\tau_i$, and
$p_i$ ($i=1,2$) are the height, characteristic time scale, and slope
of a Harvey-like profile. Further details are given by
\citet{Kamiaka+2018}.

The most readily observed multiplets are the
dipole ($\ell=1$) and quadrupole ($\ell=2$) modes, for which
\begin{eqnarray}
  \label{eq:visibility}
  & \displaystyle
  \mathcal{E}_{1,0} = \cos^2 \is,
  ~~\mathcal{E}_{1,\pm 1} = \frac{1}{2}\sin^2 \is, \nonumber \\
  & \displaystyle
  \mathcal{E}_{2,0} = \frac{1}{4} (3\cos^2 \is - 1)^2,~~
  \mathcal{E}_{2,\pm 1} = \frac{3}{8} \sin^2 2 \is,~~
  \mathcal{E}_{2,\pm 2} = \frac{3}{8} \sin^4 \is.
\end{eqnarray}
For a star with $\is = 90^\circ$, the central peak ($m=0$) is missing,
while for $\is = 0^\circ$, only the central peak is visible
\citep[e.g.,][]{Gizon2003}.

The power spectrum was analyzed using a Markov Chain Monte Carlo
algorithm based on a Metropolis-Hasting scheme, with parallel
tempering. We divided the analysis into three steps: the burn-in
phase, training phase, and acquire phase.  The burn-in phase (40{,}000
samples) ensures that we reach the region of interest in the parameter
space.  The training phase (700{,}000 samples) employs an adaptive
algorithm to optimize the covariance matrix of the Gaussian proposal
probability density function to achieve the ideal acceptance rate of
23.4\% \citep{Atchade2006}.  During the acquire phase ($10^6$
samples), the optimal covariance matrix is used to sample the
posterior distribution. Convergence of the posterior distribution is
confirmed through the Heidelberg-Welch and Geweke tests.

\begin{figure}[htb]
\centering
\rotatebox{90}{\includegraphics[width=9cm]{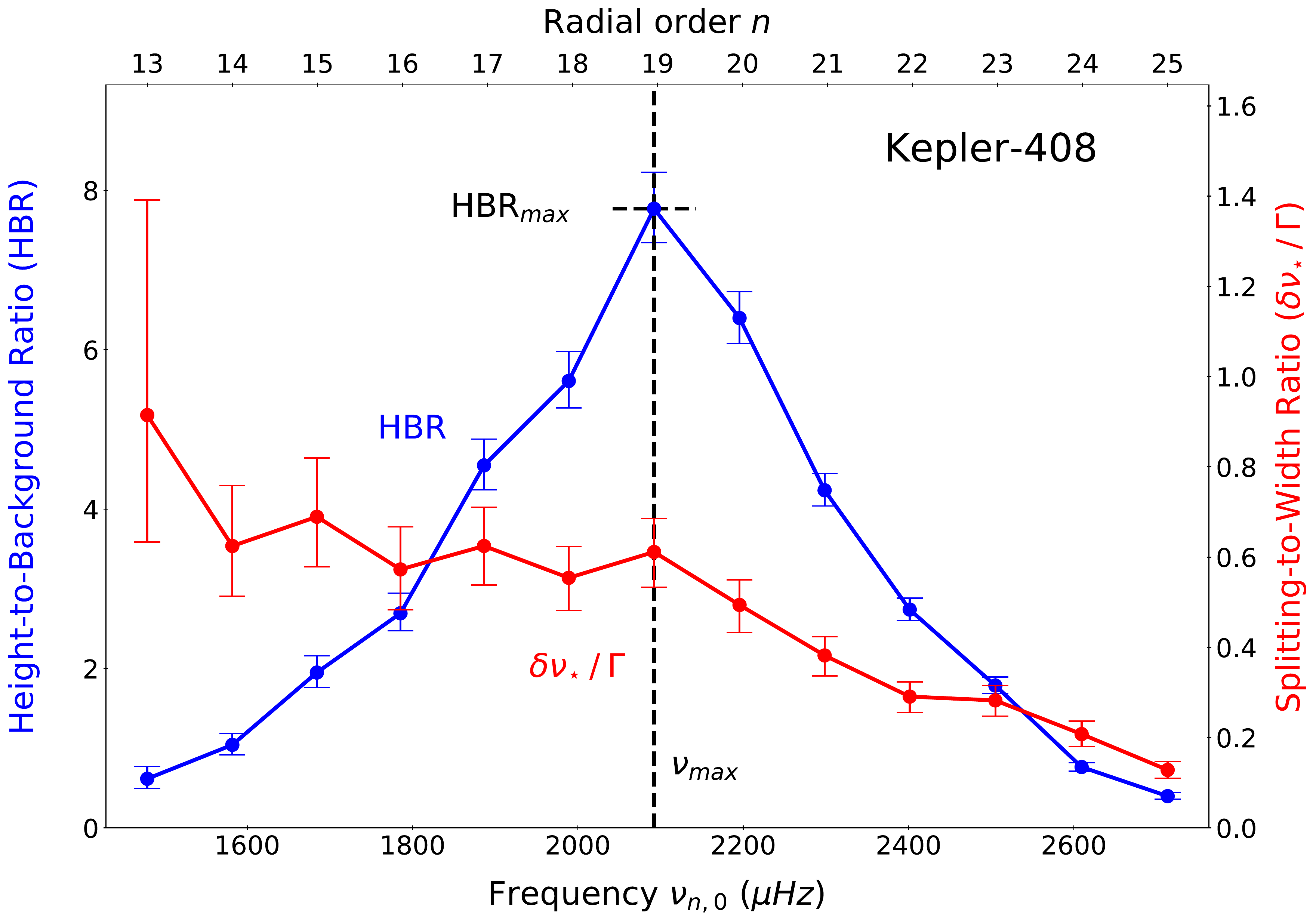}}
\caption{ Key parameters for the reliable inference of rotational
  inclination.  The blue curve shows the height-to-background ratio
  (HBR) for the $\ell=0$ modes.  The red curve shows the ratio
  between the frequency splitting $\spl$ and the line width $\Gamma$
  for the $\ell=0$ modes.}
\label{fig:408profile}
\end{figure}

Because of the large number of free parameters, we fitted the spectrum
in two steps. First, we concentrated on fitting the background, using
a single Gaussian function to model the envelope of excess power from
the oscillation modes.  The results for the background model were then
used as priors when fitting for the parameters of the oscillation
modes.  Figure~\ref{fig:408profile} plots the height-to-background
ratio (HBR) and the splitting-to-width ratio
($\delta\nu_{\star}/\Gamma$) as a function of mode frequency
$\nu_{n,l=0}$, showing that HBR takes the maximum value ${\rm
  HBR}_{\rm max}$ at $\nu_{\rm max}$.  \citet{Kamiaka+2018} found that
reliable inference of $\is$ practically requires at least ${\rm
  HBR}_{\rm max}\,{\gtrsim}\,1$ and $\delta\nu_{\star}/\Gamma(\nu_{\rm
  max})\,{\gtrsim}\,1/2$.  As indicated by the dashed line in
Figure~\ref{fig:408profile}, their criteria are satisfied for
Kepler-408.
  
\begin{figure}[ht]
\centering
\includegraphics[width=0.7\columnwidth]{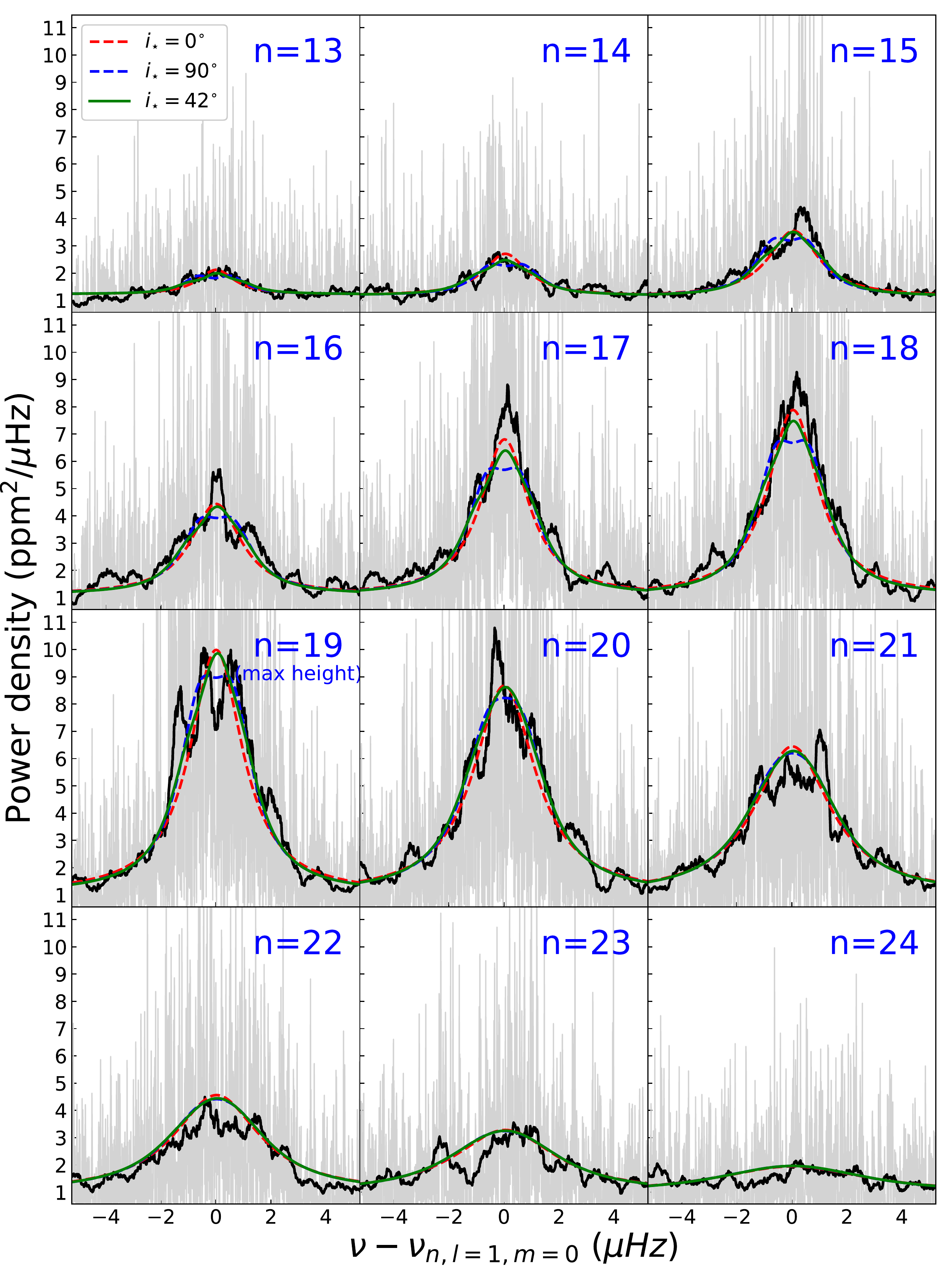}
\caption{ Individual profiles for dipole modes ($\ell = 1$) for the
  radial orders from $n=13$ to $24$.  In each panel, the gray and
  black lines represent the unsmoothed and smoothed (with a boxcar
  kernel of width 0.5\,$\mu$Hz), respectively.  The fitting results
  assuming $\is = 0^{\circ}$ and $90^{\circ}$ are plotted in red and
  blue, respectively, while our best-fit model with treating $\is$ as
  a free parameter is plotted in green ($\is = 42^{\circ}$).}
\label{fig:l=1}
\end{figure}

\begin{figure}[ht]
\centering
\includegraphics[width=0.7\columnwidth]{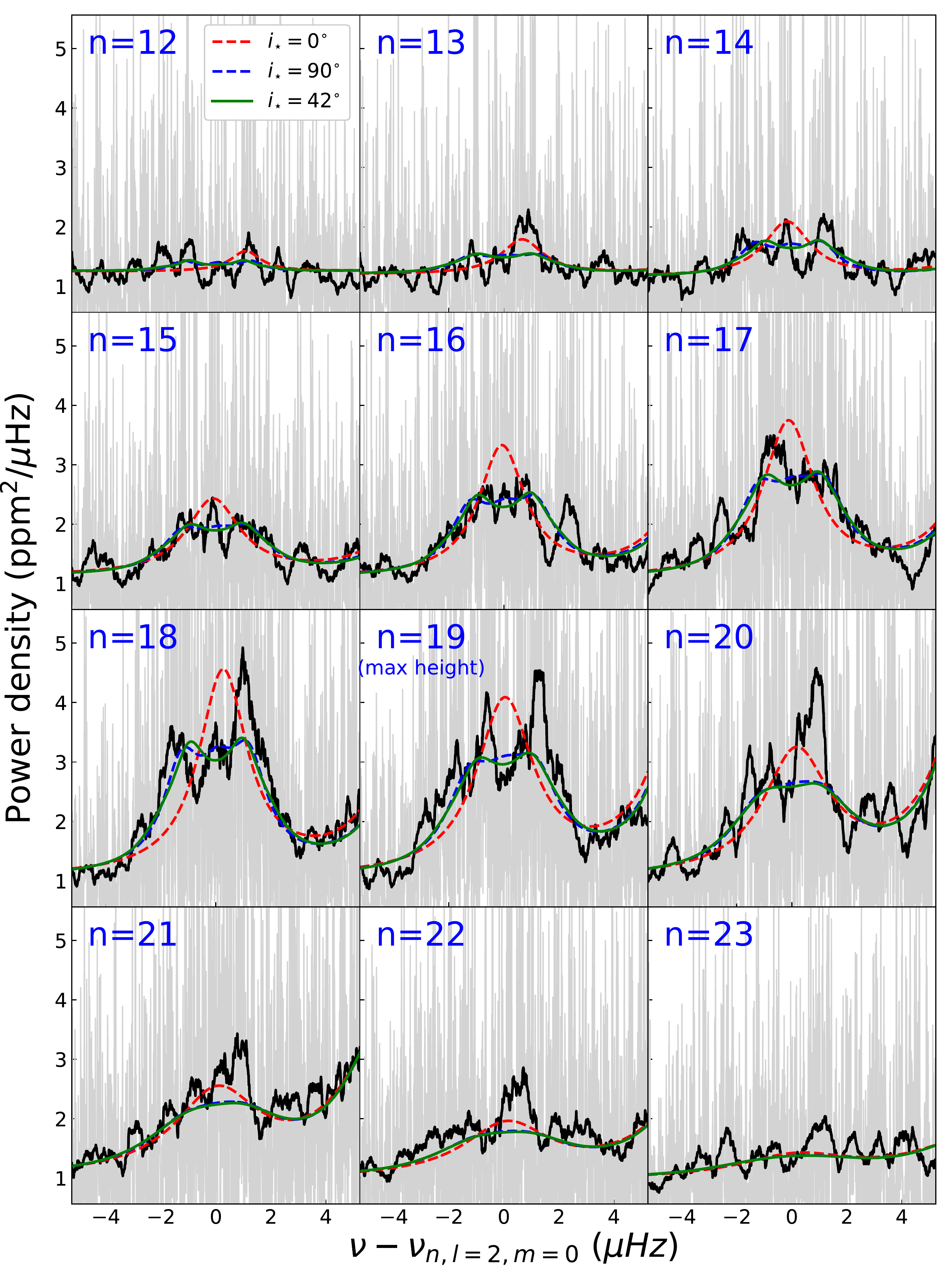}
\caption{Same as Figure \ref{fig:l=1}, but for quadrupole modes
  ($\ell=2$) from $n=12$ to $23$.}
\label{fig:l=2}
\end{figure}

We inspected each line-profile visually so as to avoid too noisy
modes, and selected the radial orders of $13 \leq n \leq 25$ for
$\ell=0$ and $1$, and $12 \leq n \leq 24$ for $\ell=2$, respectively,
for the analysis.  Figures~\ref{fig:l=1} and \ref{fig:l=2} give the
mode profiles for $\ell=1$ ($13 \leq n \leq 24$) and for $\ell=2$ ($12
\leq n \leq 23$), respectively.  In those panels our best-fits of $\is
= 42^{+5}_{-4}$ degrees and $\spl = 0.99\pm 0.10~\mu$Hz are plotted in
solid green lines (see Figure~\ref{fig:correlation} below).

\begin{figure}[ht]
\centering
\includegraphics[width=\columnwidth]{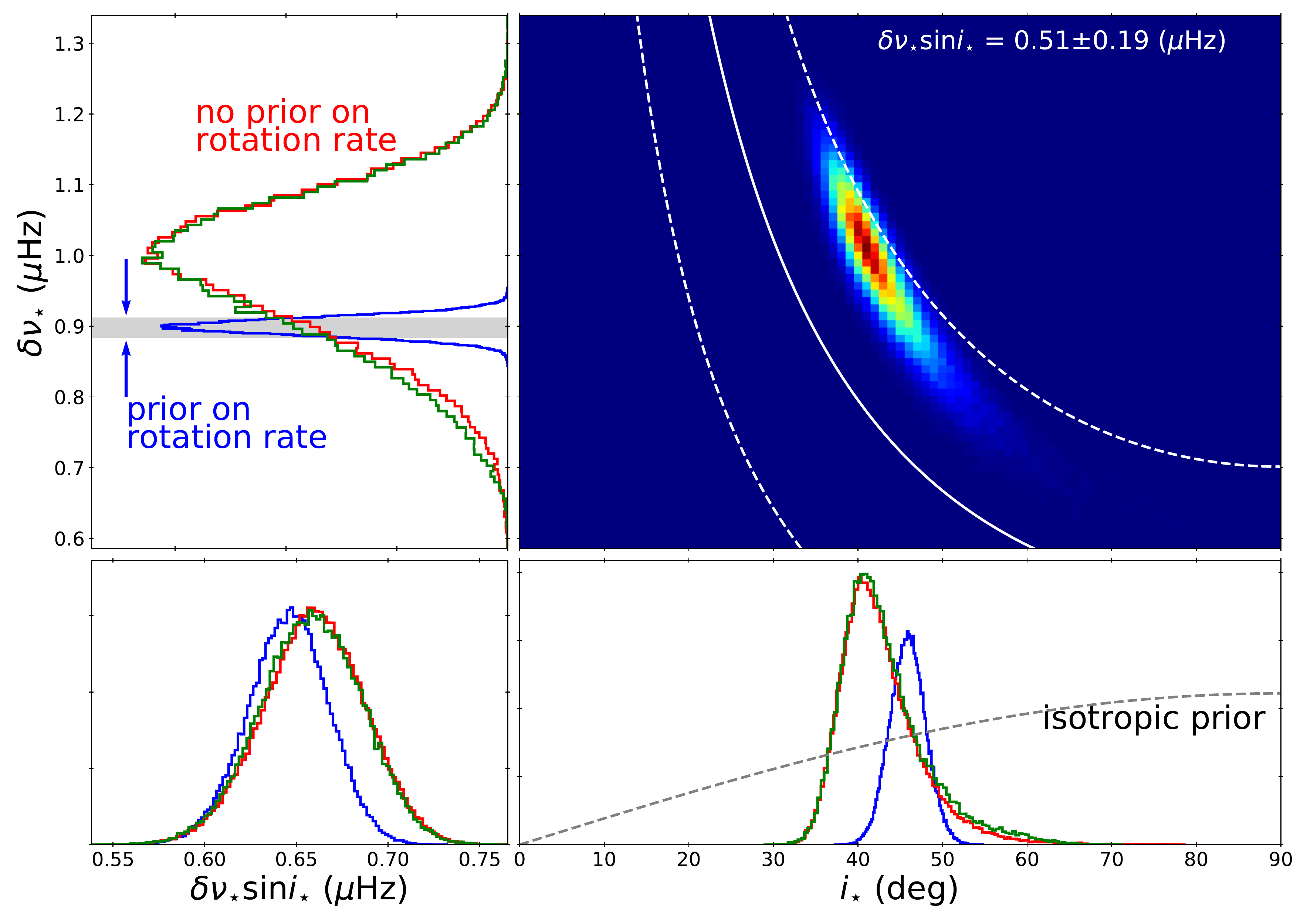}
\caption{ Constraints on rotational inclination and frequency
  splitting.  Shown is the posterior probability density (PPD) in the
  space of $\is$ and $\spl$, marginalized over all other parameters.
  The one-dimensional marginalized densities are also shown to the
  left and below the axes.  The panel in the bottom left is the PPD of
  $\spl\sin{\is}$, which is more tightly constrained than either
  $\spl$ or $\is$.  The red and blue histograms are the PPDs without
  and with a prior constraint of $\spl = 0.898\pm 0.013~\mu$Hz based
  on the measured rotation period.  The white lines identify the
  region where $\spl \sin\is=0.51 \pm 0.19\,\mu{\rm Hz}$, the value
  that is independently determined from measurements of $\vsini$ and
  $R_\star$ (see Section \ref{sec:spectroscopy}).  The
    bottom-right panel indicates the marginalized PPD for the stellar
    inclination $\is$.  The median and 68\% (95\%) credible interval
    are $\is=41.7_{-3.5}^{+5.1}$ ($\is=41.7_{-6.4}^{+13.3}$) based on
    asteroseismology alone.  When the prior on rotation rate is
    applied, the results become $\is=45.8_{-2.2}^{+2.1}$
    ($\is=45.8_{-4.3}^{+4.4}$). The result incorporating the
    apodization factor in our background noise model (green lines) is
    almost indistinguishable (see Section 4.4). }
\label{fig:correlation}
\end{figure}

\subsection{Checks for consistency and robustness}
\label{subsec:checks}

The measured splitting is in agreement with the value of $1/P_{\rm
  rot} = 0.898\pm 0.013~\mu$Hz based on the photometric rotation
period, thereby providing a successful consistency check.  We also
tried using the photometric rotation period as a prior constraint on
the asteroseismic analysis, which sharpened the constraint on the
stellar inclination angle to $45.9{\pm}2.1$ degrees (see the blue
curves in Figure~\ref{fig:correlation}).

\begin{figure}[ht]
\centering
\includegraphics[width=0.70\columnwidth]{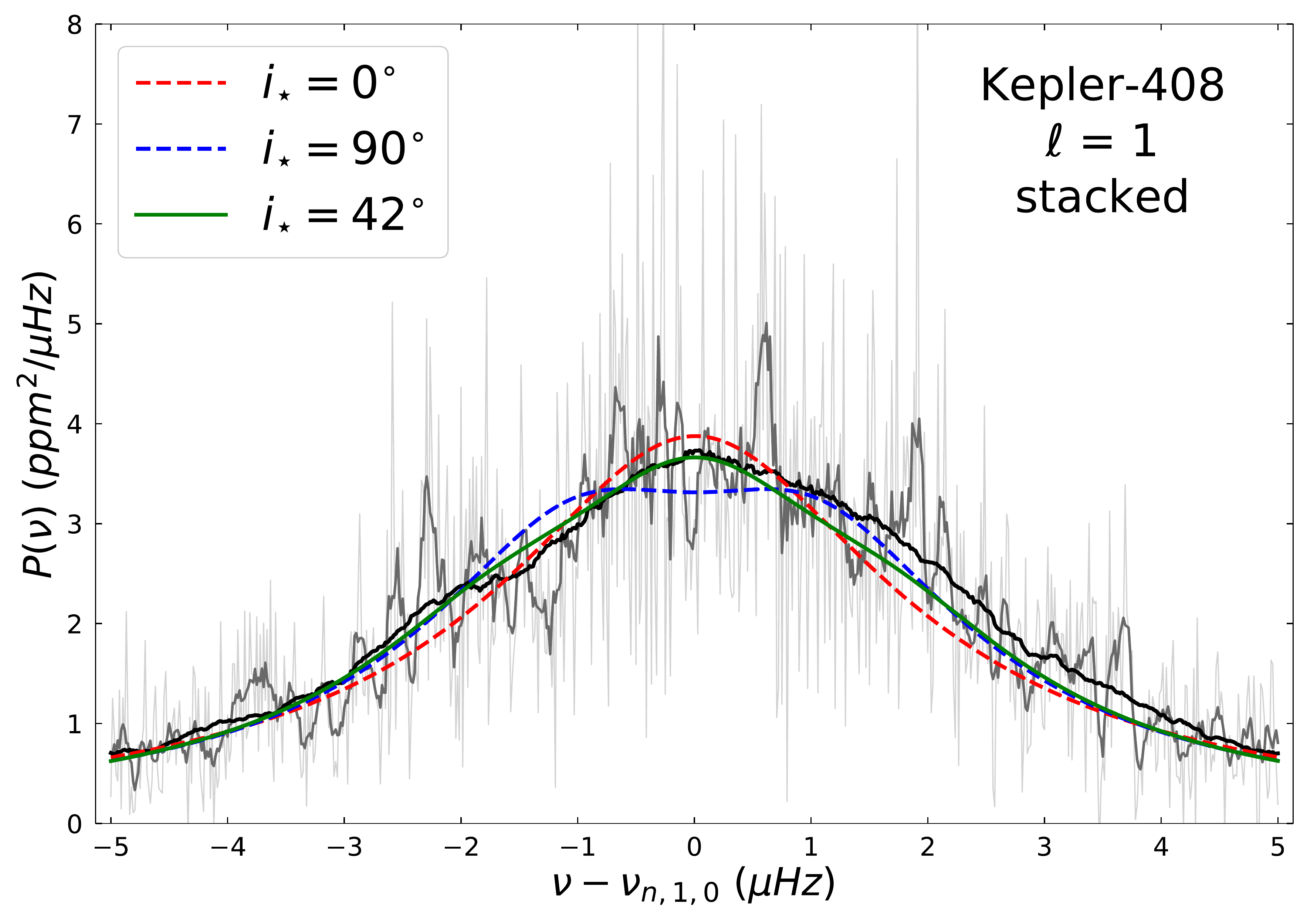}
\includegraphics[width=0.70\columnwidth]{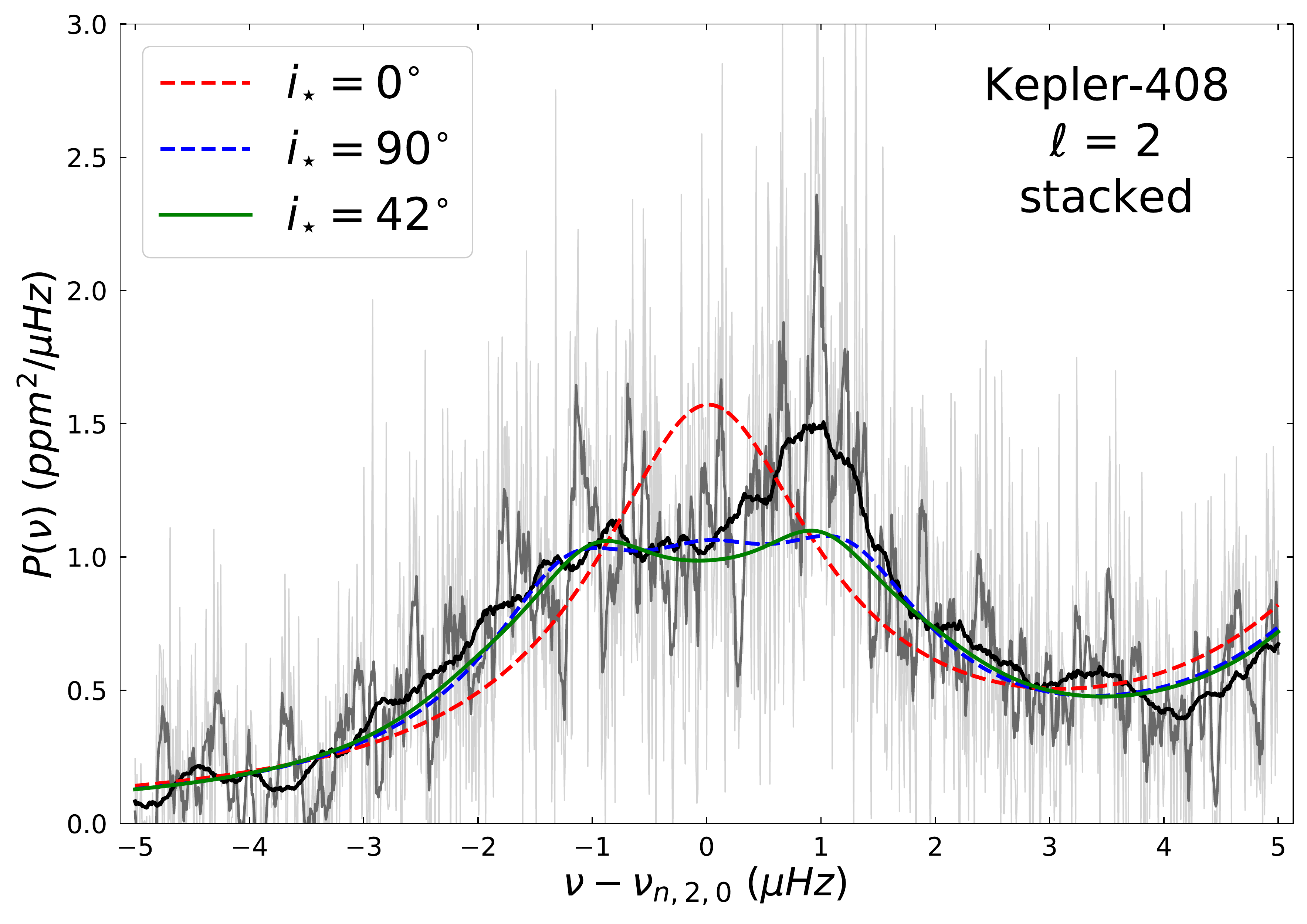}
\caption{ Average power spectra of rotationally split multiplets, for
  $\ell=1$ (top) and $\ell=2$ (bottom).  The profiles of multiple
  modes have been stacked to improve the signal-to-noise ratio and
  allow for a visual inspection, although the quantitative fits were
  performed on the data without any averaging or stacking (see
  Figs.~\ref{fig:l=1} and \ref{fig:l=2}).  For $\ell=1$, the modes
  with $n=13$ to $25$ were included.  For $\ell=2$, the modes with
  $n=12$ to $24$ were included.  The thin gray line shows the data
  without any smoothing, while the thick gray and black lines show the
  data after smoothing over 0.05 and 0.75\,$\mu$Hz in frequency,
  respectively.  Each panel also shows three model curves that were
  optimized to fit the data.  The red curve is based on a model
  assuming $\is = 0^{\circ}$, the blue curve is for $\is =
  90^{\circ}$, and the green curve is for a model in which $\is$ is a
  free parameter.  For the $\ell=2$ modes, the gradual rise observed
  at the high-frequency end is from a neighboring radial mode
  ($\ell=0$). The asymmetry in the line profile is not understood (see
  Section~\ref{subsec:checks}).}
\label{fig:stacked}
\end{figure}

To allow for a visual inspection, Figure~\ref{fig:stacked} displays
the average $\ell=1$ and $\ell=2$ profiles, based on the combination
of the data from 13 different radial orders.  The profile of the
average $\ell=1$ multiplet (the top panel) is centrally peaked,
demonstrating the visibility of the $m=0$ mode, and ruling out an
inclination angle near $90^\circ$.  The signal-to-noise ratio and
frequency resolution are high enough that the absence of the $m=0$
mode would have led to a flat-topped appearance, from the combination
of the marginally resolved $m=+1$ and $-1$ modes.  On the other hand,
the profile of the $\ell=2$ modes is not centrally peaked, ruling out
inclinations near zero.  Together, the appearance of the modes suggests
an intermediate value of the inclination. 

\begin{figure}[ht!]
\centering
\includegraphics[width=0.48\columnwidth]{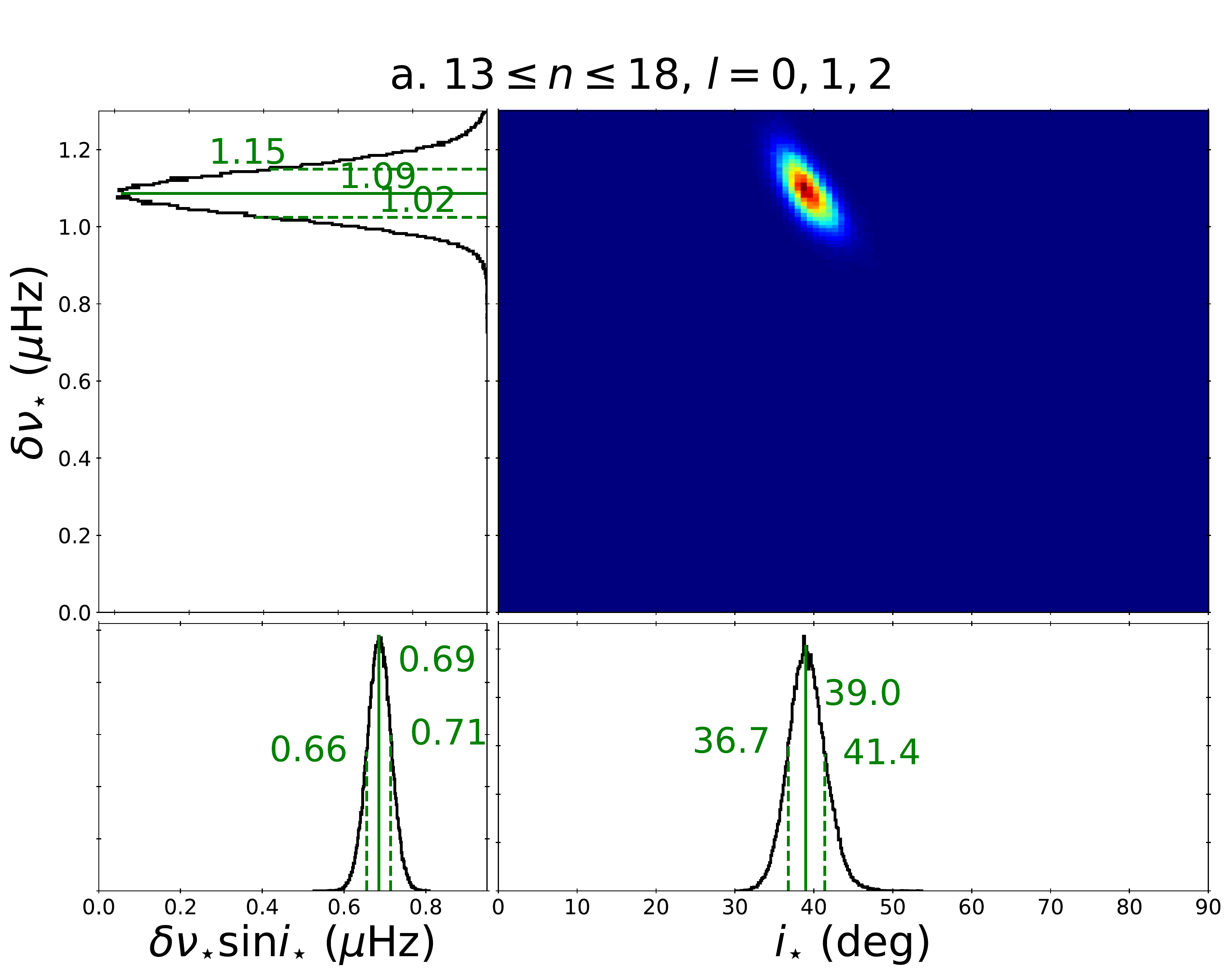}
\includegraphics[width=0.48\columnwidth]{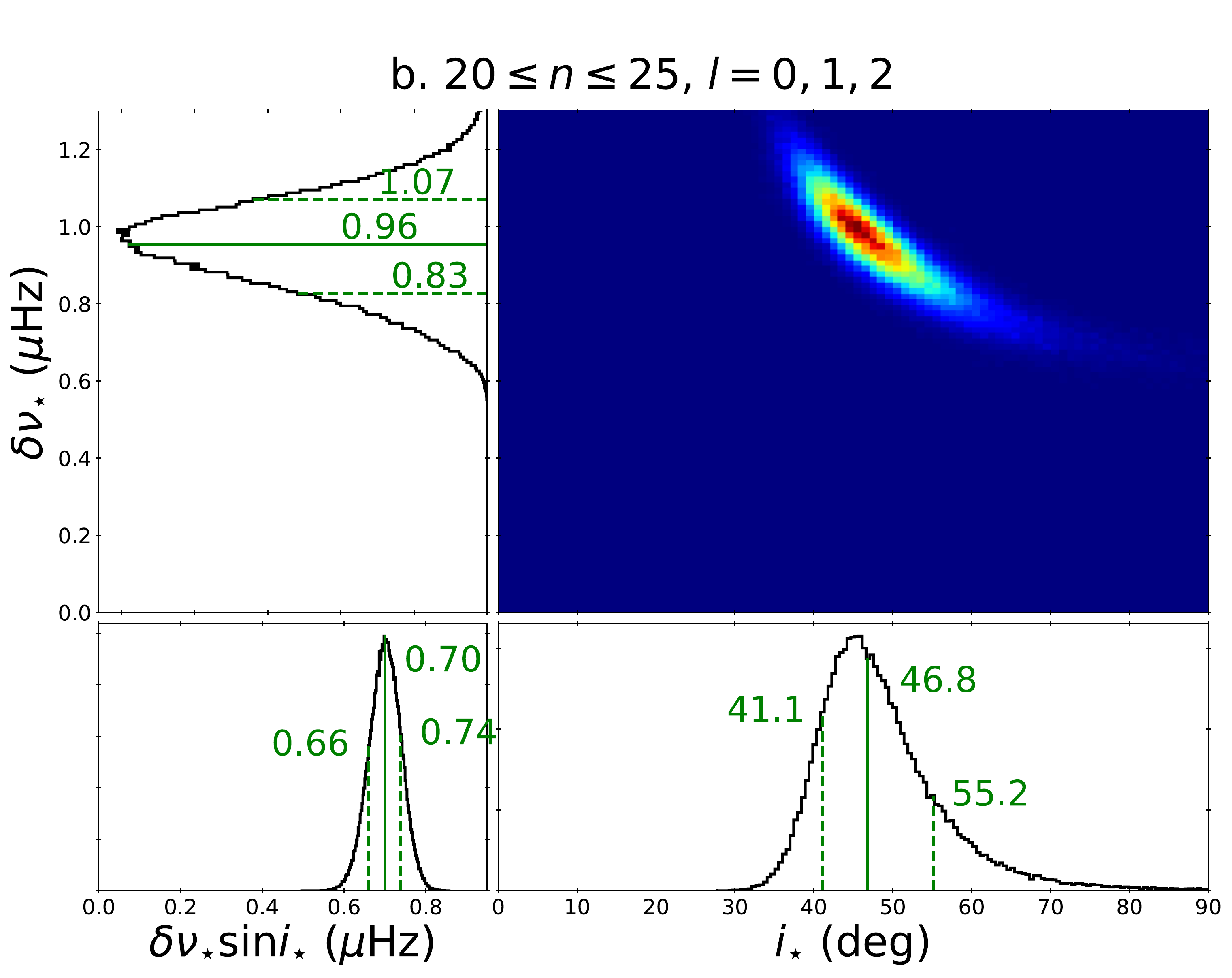}\\
\includegraphics[width=0.48\columnwidth]{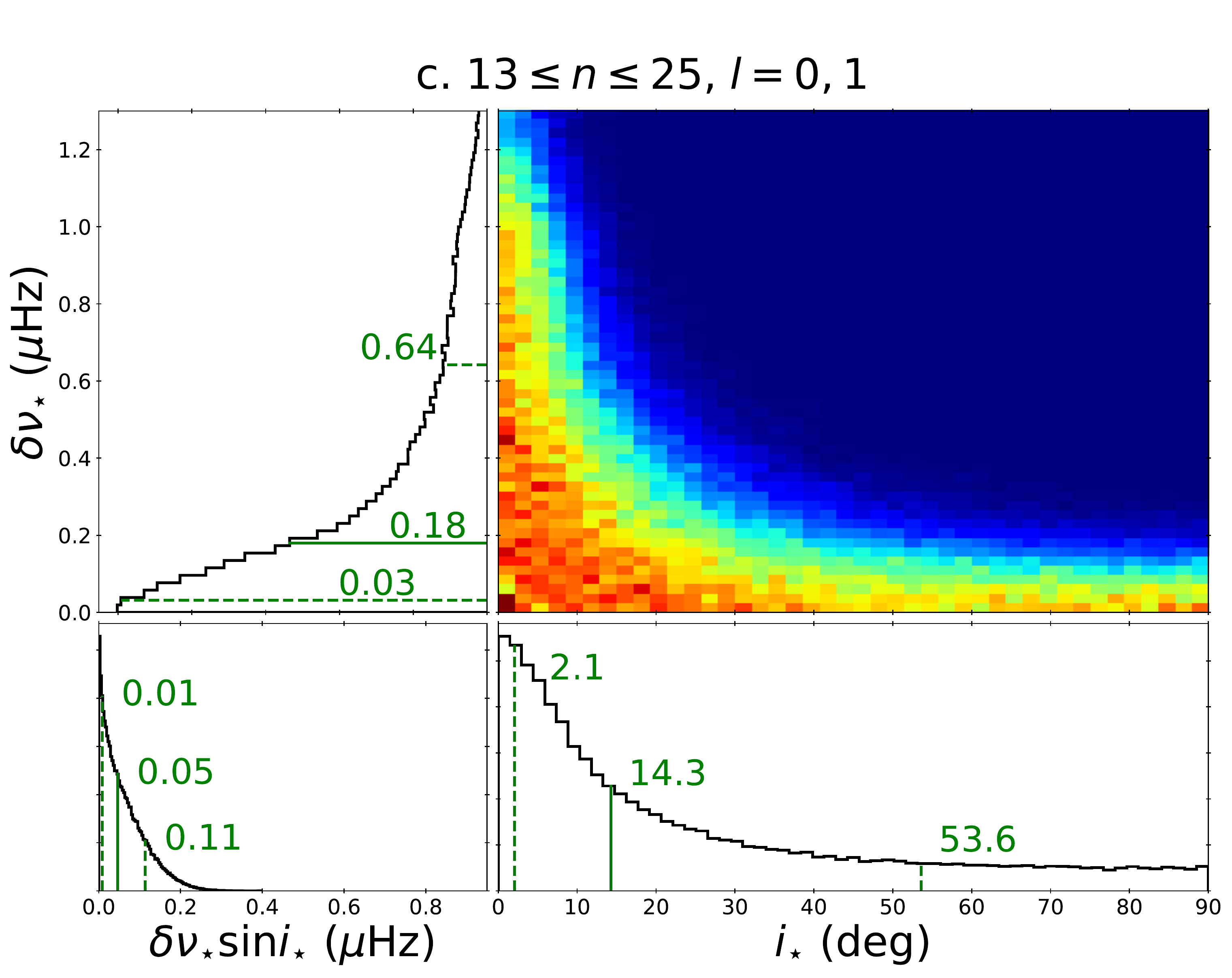}
\includegraphics[width=0.48\columnwidth]{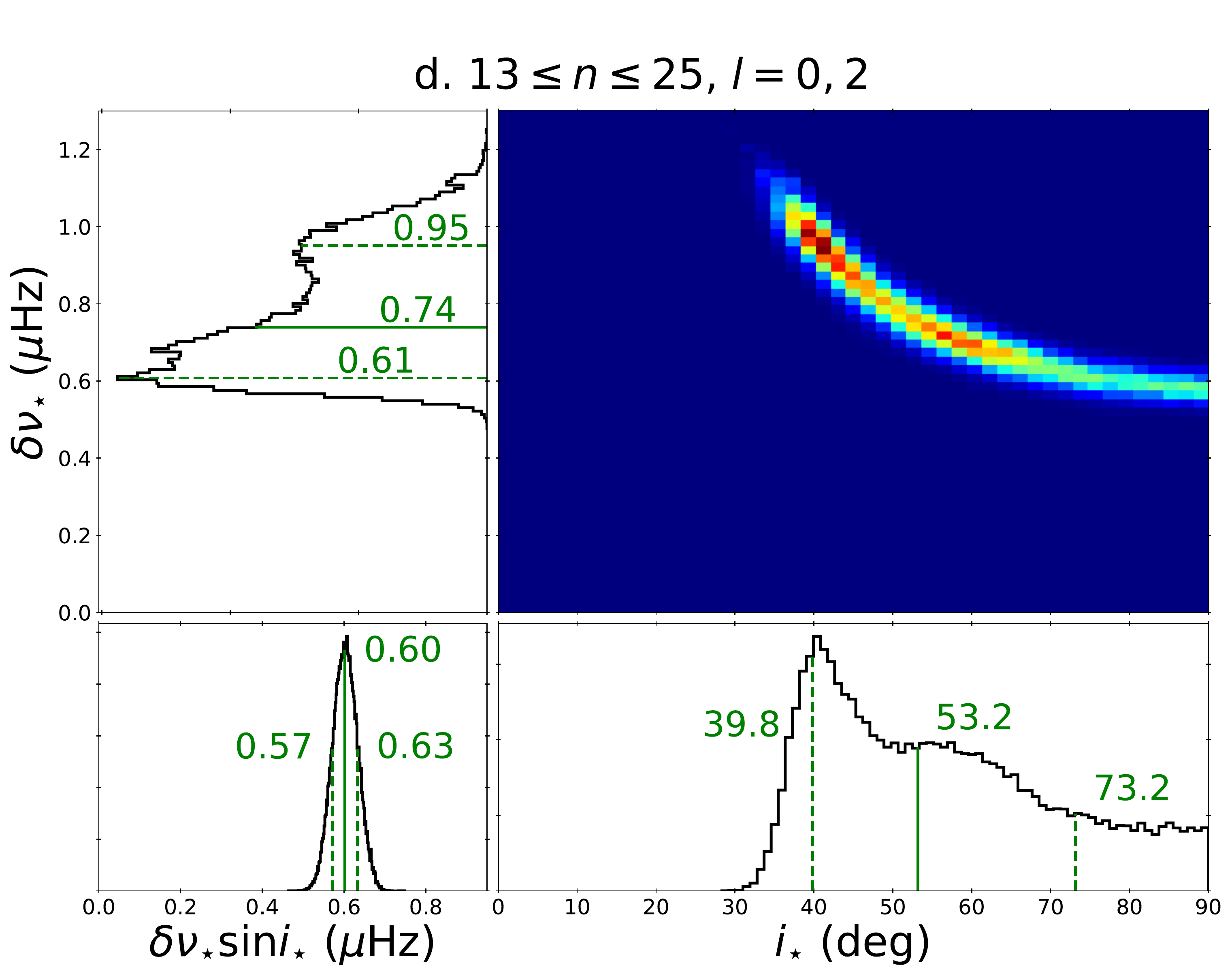}\\
\includegraphics[width=0.48\columnwidth]{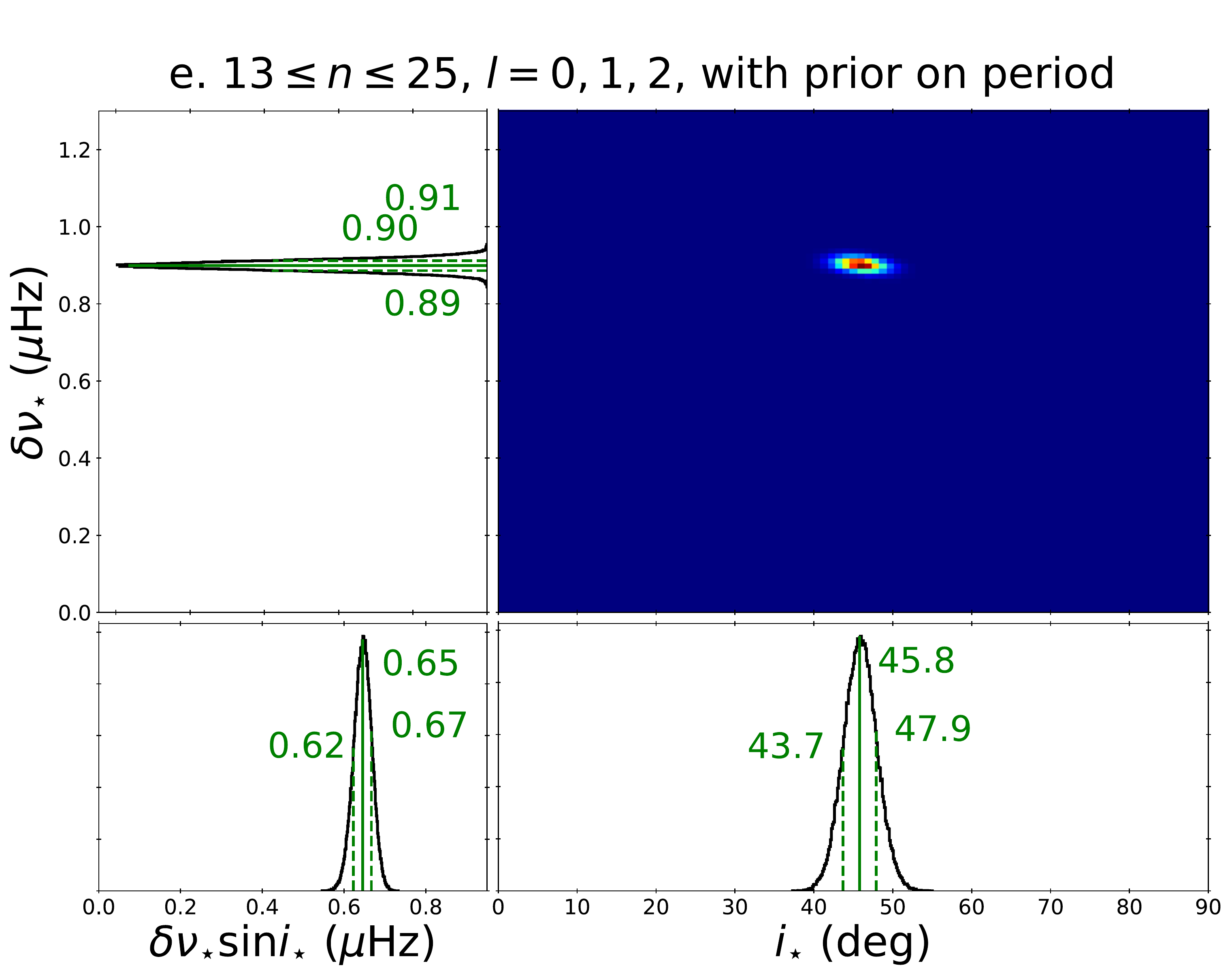}
\includegraphics[width=0.48\columnwidth]{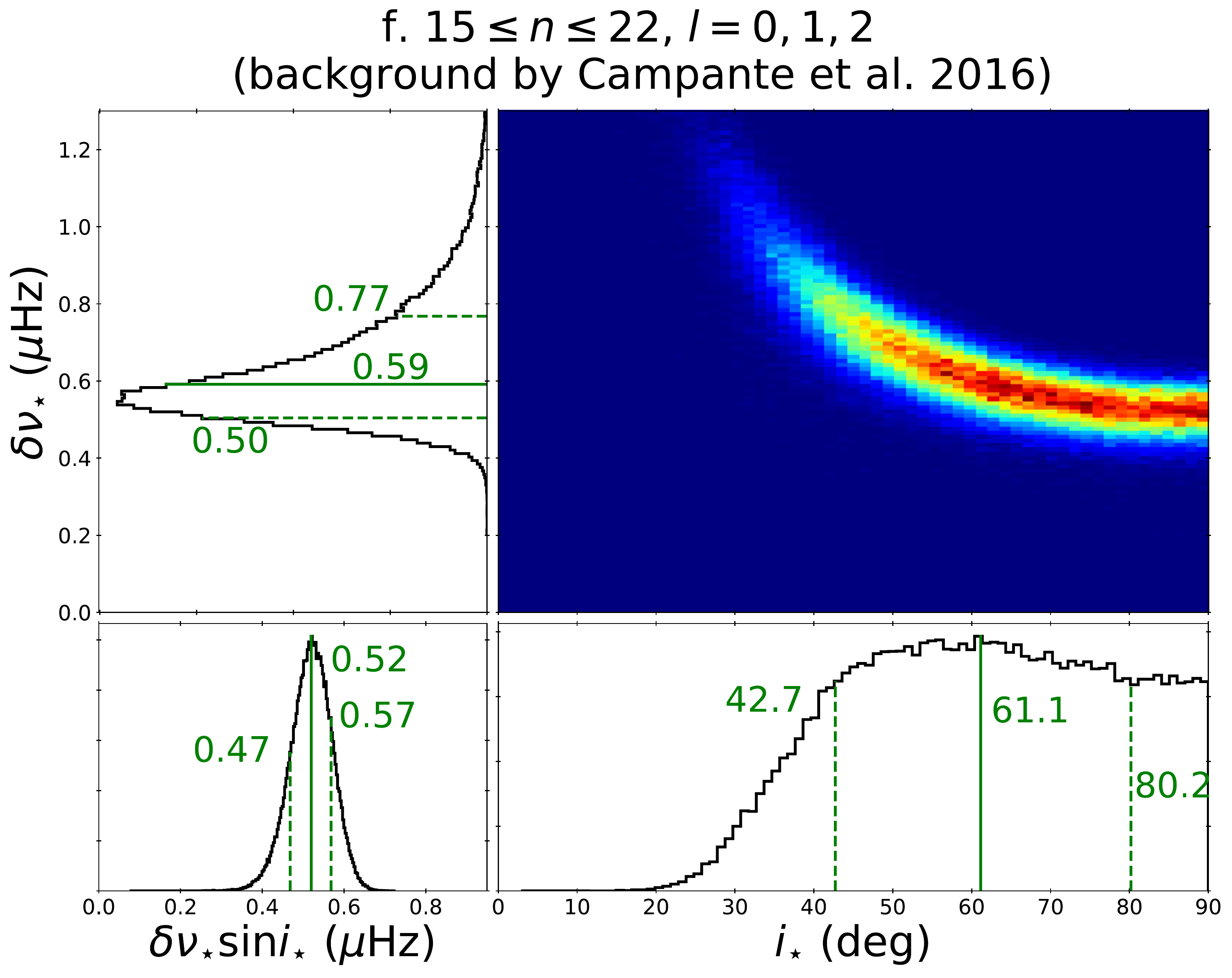}\\
\caption{ Constraints on $\is$ and $\spl$, as in
  Fig.~\ref{fig:correlation}, after making variations in the analysis
  procedure.  {\it a}: Fitting only the lower radial orders ($13 \leq
  n \leq 18$) and $\ell=0,1,2$.  {\it b}: Fitting only the higher
  radial orders ($20 \leq n \leq 25$) and $\ell=0,1,2$.  {\it c}:
  Fitting only the radial and dipole modes ($\ell=0,1$) of all orders.
  {\it d}: Fitting only the radial and quadrupole modes ($\ell=0,2$)
  of all orders.  {\it e}: Fitting $\ell=0,1,2$ of all orders, with a
  Gaussian prior of $\spl = 0.898\pm\ 0.013~\mu$Hz based on the
  measured rotation period \citep{Angus+2018}.  {\it f}: Fitting all
  orders and modes, after replacing our model for the noise background
  with the (unsatisfactory) model of \citet{Campante+2016}. The green
  solid and dashed lines in the histograms indicate the median and
  $1\sigma$ credible regions.  
  }
\label{fig:correlation6}
\end{figure}

The bottom panel of Figure~\ref{fig:stacked} shows that the $\ell=2$
multiplet has an asymmetric appearance, with more power at frequencies
above the line center than below.  This is unexpected because the
geometrical factors $\mathcal{E}_{\ell,m}$ do not depend on the sign
of $m$.  Figure \ref{fig:l=2} suggests that this asymmetry in power is
mainly due to modes of high radial order ($n=18$ to $21$). Such
high-order modes are more sensitive to the conditions near the stellar
surface \citep[e.g.][]{JCD1997, Kjeldsen2008, Ball2014,
  Sonoi2015}. Thus, the observed asymmetry may arise from the (poorly
understood) magnetic and non-adiabatic processes occurring near the
surface.  We performed the similar stacking analysis for several stars
in \citet{Kamiaka+2018}, all of which do not exhibit any noticeable
asymmetry. Thus the asymmetry seems fairly specific to Kepler-408.
Because the reason for the asymmetry is not clear, we tried fitting
only the $\ell=1$ modes and found that a low inclination (and a high
stellar obliquity) are still preferred as shown in
Figure~\ref{fig:correlation6}c below.

We repeated the fit after decreasing the amplitude of the (unsmoothed)
data by 30 \% for $-\Gamma_{n,l=2,m=0}/2 < \nu -\nu_{n,l=2,m=1}<
\Gamma_{n,l=2,m=0}/2$ for $18 \leq n \leq 21$, and found that $\is
=39^{+4}_{-3}$ degrees. Since fitting these simulated spectra did not
lead to any systematic bias in the results for the inclination, the
observed asymmetry for the $\ell=2$ modes does not affect our
conclusion that Kepler-408b has a significantly misaligned orbit.

As further tests of robustness, we repeated the analysis for 5
different choices of the set of radial orders and angular degrees to
be fitted (see Figure~\ref{fig:correlation6}).  This led to larger
uncertainties, and small systematic changes in the derived parameters.
Fitting the $\ell=1$ modes tends to give lower inclinations, while the
$\ell=2$ modes favor higher inclinations. Such complementary
roles of $\ell=1$ and $\ell=2$ modes are very useful in constraining $\is$
and $\spl$ reliably. While Kepler-408 is one of the stars with the
clearest pulsation spectrum, its asteroseismic modeling is still
subtle and careful individual tests are required for the reliable
parameter extraction.

In all cases, though, the results are compatible with a large
spin-orbit misalignment, and the splitting is compatible with the
photometric rotation period, implying that our asteroseismic
inference for the Kepler-408 system is robust.

\subsection{Comparison with previous results}

In reality, our results do not agree with those of
\citet{Campante+2016}, who found $\is>54^\circ$ within $1\sigma$
confidence (we note, however, that their 95.4\% constraint is
  $\is>36.5^\circ$).  In attempting to understand the reason for the
  discrepancy, we realized that we use an unweighted power spectrum
  (Lomb-Scargle periodogram), while \citet{Campante+2016} computed it
  using a weighted least-square-fitting method. Thus we repeated our
  analysis using their spectrum, and obtained almost the same
  inclination angle, implying that the difference of the spectra is
  not a major reason for the discrepancy.  We also noticed that their
best fitting model gave $\spl = 0.50_{-0.04}^{+0.20}~\mu$Hz, which is
inconsistent with the photometrically measured rotation period.
Another difference is related to the chosen model for the background
noise in the power spectrum.  For the sake of uniformity,
\cite{Campante+2016} adopted the same model for all 25 systems of
their analysis. Their model was parameterized as:
\begin{eqnarray}
\label{eq:noise-campante}
  B(\nu) = B_{0} +
  \left[\frac{B_{1}}{1+(2\pi{\nu}\tau_1)^{a}} + \frac{B_{2}}{\nu^{2}}\right]
  {\rm sinc}^2 \left( \frac{\pi \nu }{2 \nu_0} \right),
\end{eqnarray}
where $\nu_0 = 8496.6~\mu$Hz is the Nyquist frequency.

\begin{figure}[ht]
\centering
\includegraphics[height=5cm]{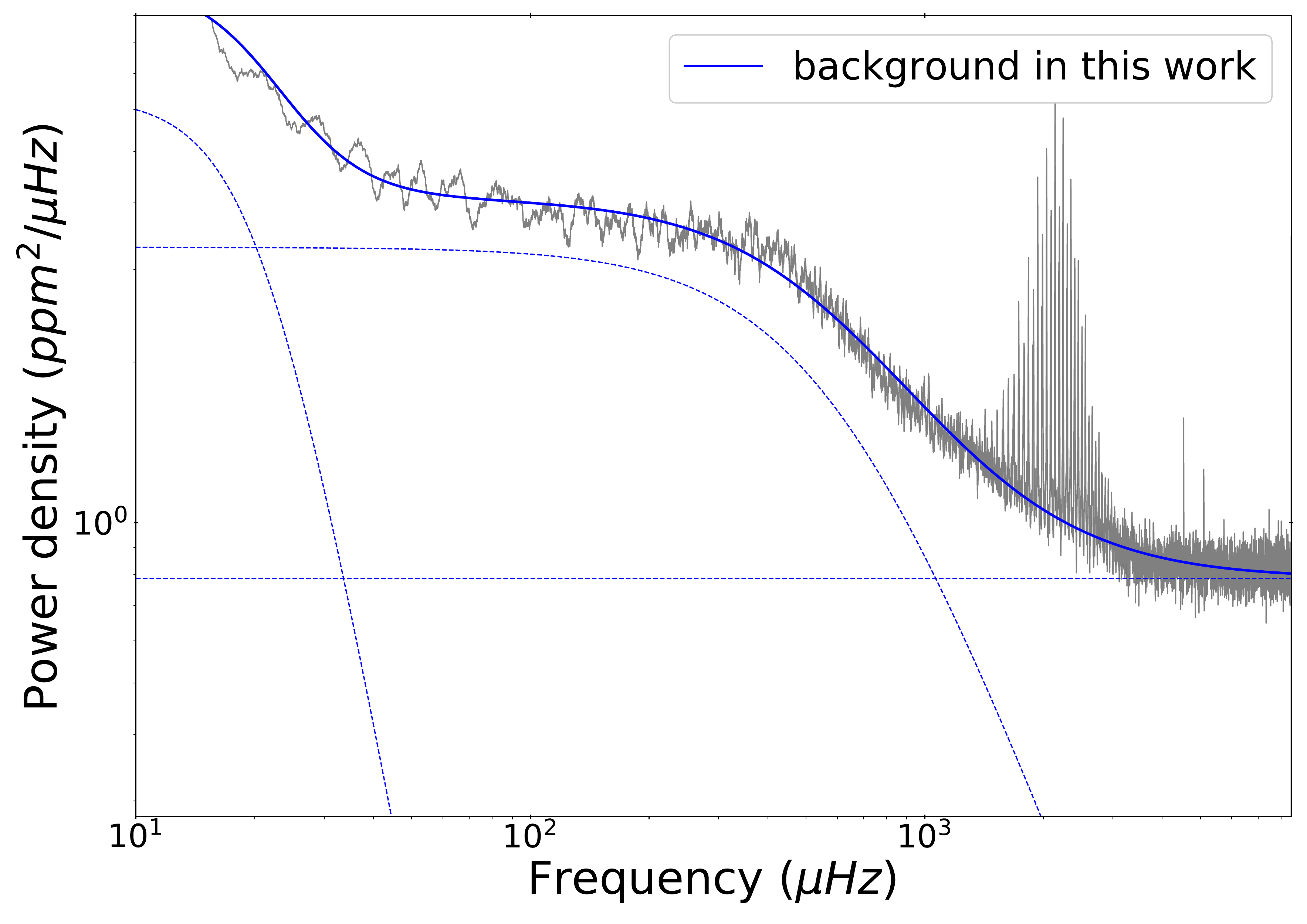}
\includegraphics[height=5cm]{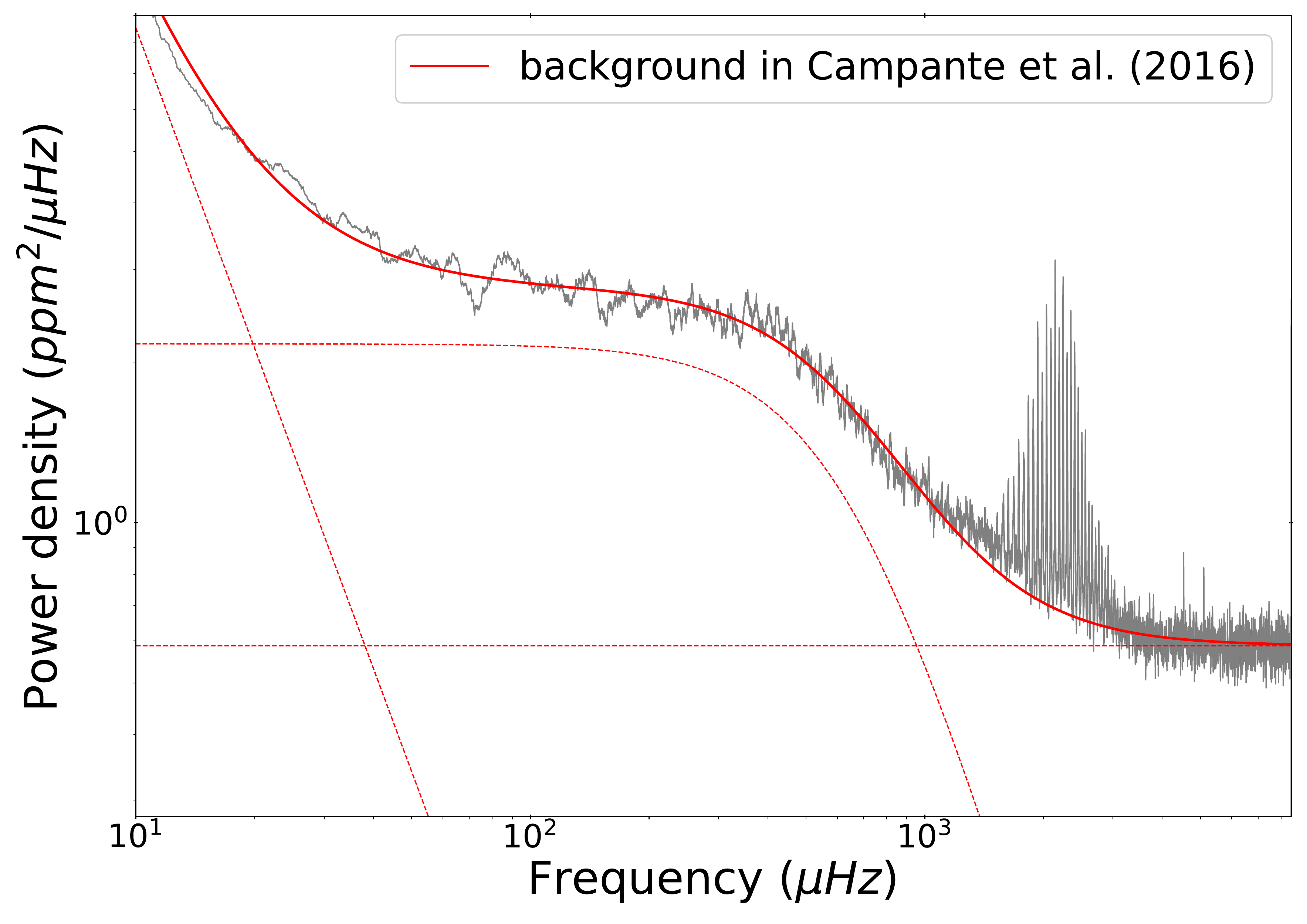}\\
\rotatebox{90}{\includegraphics[width=5cm]{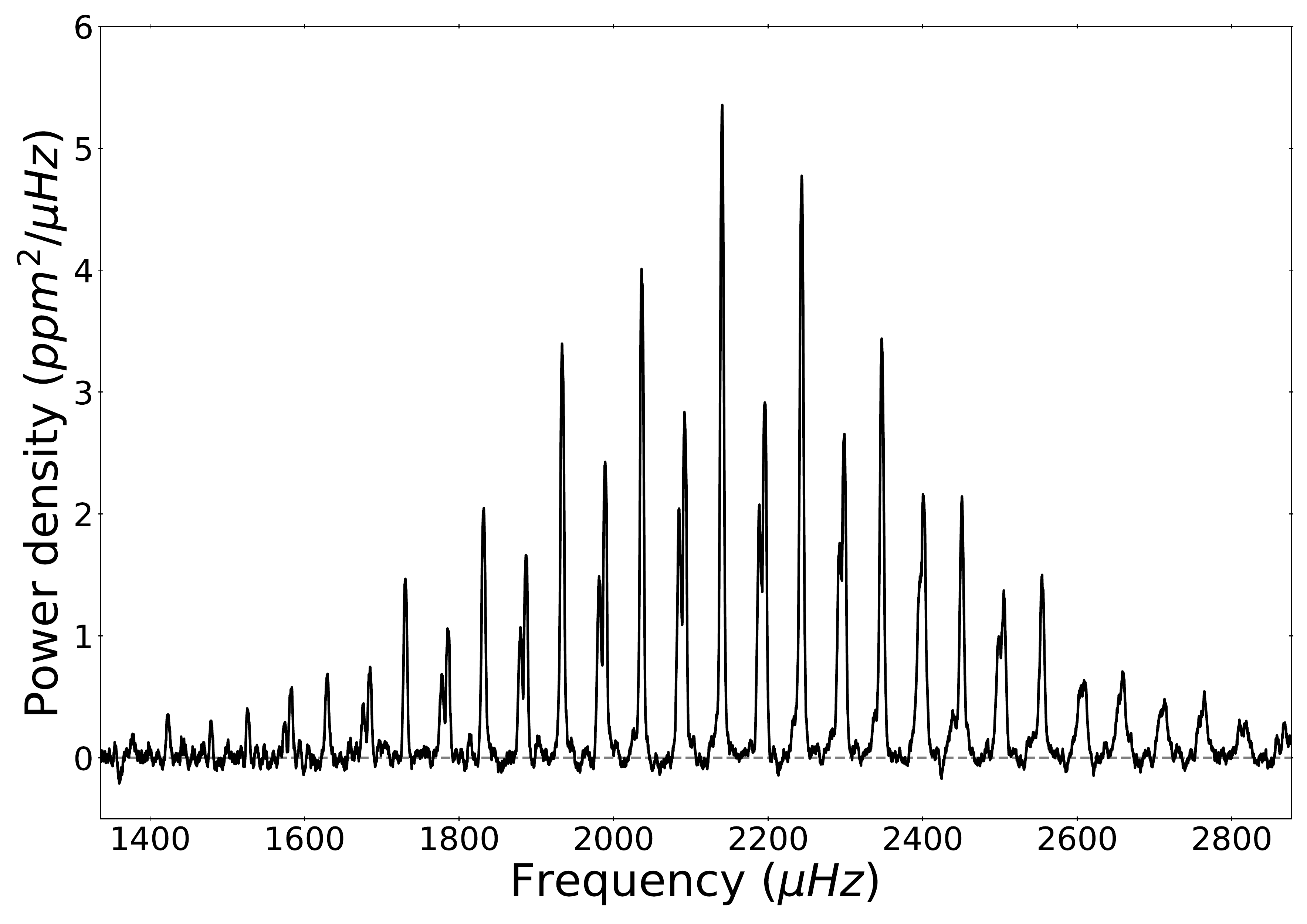}}
\rotatebox{90}{\includegraphics[width=5cm]{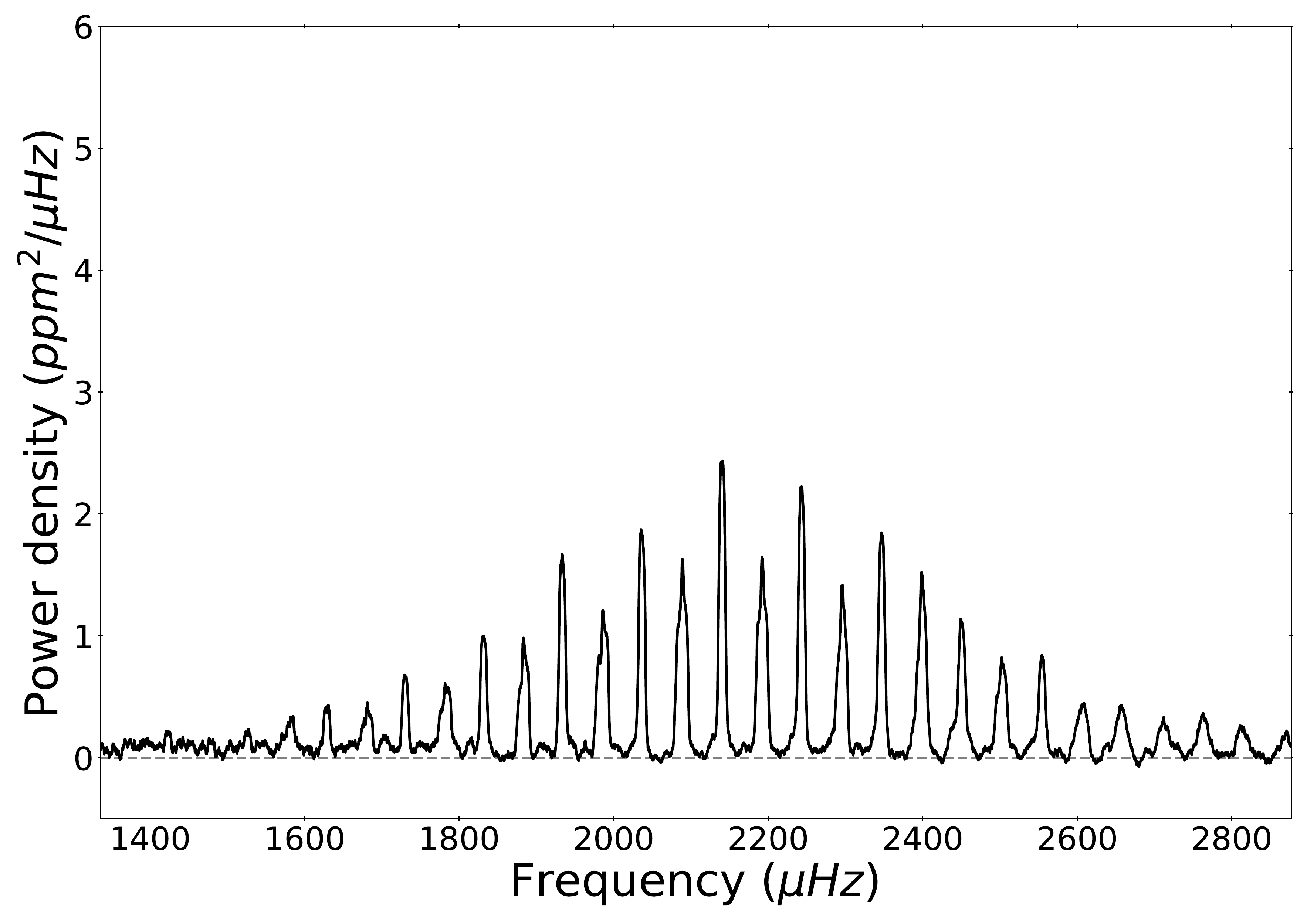}}
\caption{ Models of the noise background.  {\it Top panels.} --- The
  entire power spectrum of Kepler-408, along with the best-fitting
  model and its three separate components.  The left panel shows the
  spectrum and model used in our analysis.  The right panel shows
  those used in \citet{Campante+2016}, which does not fit well
  the lower envelope of the power spectrum in the vicinity of the
  oscillation modes.  {\it Bottom panels.} --- Close-up of the
  oscillation modes, after subtracting the best-fitting model for the
  background.  }
\label{fig:408noisebackground}
\end{figure}

While equation (\ref{eq:noise-campante}) works reasonably well in
general, the residuals from the best-fit of their noise model (right
panels of Figure~\ref{fig:408noisebackground}) shows that it poorly
fits the low-frequency part of the noise background of Kepler-408;
There is a systematic departure from the zero baseline of up to 10\%
in the vicinity of the low-frequency modes, suggesting that their
noise background fit is not satisfactory. In contrast, our background
model, equation (\ref{eq:noise-model}), fits much better, as
illustrated in the left panels of Figure~\ref{fig:408noisebackground}.
More specifically, our best-fit model based on our power spectrum with
the background model (\ref{eq:noise-model}) is preferred over that
based on the Campante et al. power spectrum with equation
(\ref{eq:noise-campante}) by an ``odds ratio'' significantly larger
than $100$.  The odds ratio is the ratio of the two integrated
likelihoods; see section 12.7 of \citet{Gregory2005} for details.
According to the \cite{Jeffreys1961} classification, this is a
decisive evidence in favor of our background noise model.

Incidentally equation (\ref{eq:noise-model}) does not include the
apodization correction factor ${\rm sinc}^2 (\pi \nu/2 \nu_0)$ unlike
equation (\ref{eq:noise-campante}). This could affect
the estimate of $\is$. We assessed this possibility by applying
  the correction for the first two terms in equation
(\ref{eq:noise-model}) as in \cite{Davies2016}, and found that
the resulting fit yields $\is=41.9^{+5.7}_{-3.6}$ degrees (see green
histogram in Figure \ref{fig:correlation}). Thus we confirm that the
apodization factor does not change our conclusion.

When we replaced our model for the background with that of equation
(\ref{eq:noise-campante}), we were able to reproduce the result of
$\is > 54^\circ$ reported by \citet{Campante+2016}.  Evidently, it is
essential to perform a careful subtraction of the low-frequency noise
for each system, to obtain an unbiased estimate of $\is$ from
asteroseismic analysis. \citet{Appourchaux2012b} have studied the
systematic errors in measurements of seismic parameters caused by
inaccuracies in the model for background noise. Although they did not
examine the implications for inference of the inclination angle, they
did note that inaccuracies can greatly impact the inferred mode
heights and linewidths, which in turn may bias the measurement of the
rotational splitting and inclination.  Our work demonstrates that this
is indeed the case: systematic errors in the background model can
severely bias the measured inclination.

As an additional test, we tried replacing the background model with a
simple quadratic function of frequency. By restricting the frequency
range to the limited interval spanned by the oscillation modes
(1300--2900~$\mu$Hz), we found that the quadratic function also gives
a good fit.  The results for the inclination were the same as in
our original analysis ($\is =42^{+5}_{-4}$~degrees), confirming that
the exact functional form of the background model does not matter, as
long as it fits reasonably well.

\section{Projected rotation rate} \label{sec:spectroscopy}

There is also evidence independent of asteroseismology that the
rotational inclination is in the neighborhood of $45^\circ$, based on
the measured values of the stellar radius, rotation period, and
sky-projected rotation velocity (Table \ref{tab:parameters}).  The
stellar radius ($R_\star$) was determined by combining the observed
geometric parallax, apparent $K$ magnitude, and spectroscopic
effective temperature \citep{Berger+2018}.  The rotation period
($P_{\rm rot}$) was determined from the {\it Kepler} photometry, as
noted above.  The combination of these quantities implies $v_{\rm rot}
= 2\pi R_\star/P_{\rm rot} = 4.92 \pm 0.21$~km~s$^{-1}$.  Meanwhile,
the sky-projected rotation velocity ($\vsini$) was found to be $2.8\pm
1.0$~km~s$^{-1}$ by modeling the Doppler-rotational contribution to
the observed spectral line broadening \citep{Petigura+2017}.

Together, these data can be used to place constraints on $\sin \is$.
To obtain the likelihood function for $\sin \is$, we integrated
$p_1(v_{\rm rot}) \cdot p_2(v_{\rm rot} \sin \is)$ over $v_{\rm rot}$,
where $p_1$ and $p_2$ are Gaussian functions representing the
constraints $v_{\rm rot} = 4.92\pm 0.21$~km~s$^{-1}$ and $v_{\rm rot}
\sin\is = 2.8\pm 1.0$~km~s$^{-1}$.  The result is $\sin \is = 0.70\pm
0.21$, or $\is = 44^{+20}_{-15}$~degrees, which is consistent with our
asteroseismic result.
  
As another consistency check, we can combine the spectroscopically
determined $\vsini$ and $R_\star$ to give $\spl \sin\is=0.51 \pm 0.19~
\mu{\rm Hz}$.  The white lines in Figure \ref{fig:correlation} show
the region that is defined by this constraint, which is independent of
the asteroseismic analysis. The results are again consistent to within
1-$\sigma$.

\section{Summary} \label{sec:summary}

By modeling the power spectrum of $p$ modes, we found the stellar
inclination to be $\is=41.7_{-3.5}^{+5.1}$ ($\is=41.7_{-6.4}^{+13.3}$)~degrees with 68\% (95\%) credible interval (Section
\ref{sec:rotation}).  \citet{Nielsen+2017} and \citet{Kamiaka+2018}
previously reported a similar result for Kepler-408, but did not
remark on the conflict with the analysis of \cite{Campante+2016}, nor
did they appreciate the importance of this system for understanding
the origin of the spin-orbit misalignment (described below).  The more
thorough analysis in the present paper has resolved the conflict, by
examining the individual and stacked line-profiles for different
modes, comparing the best-fit with and without the photometric
rotation period constraint, and exploring different possibilities for
the background model.  This experience with Kepler-408 and the
methodology presented in this paper should allow for more robust
determinations of $\is$ in the future, through the precise and
accurate combination of asteroseismology, photometry and spectroscopy.

As for the inclination of the orbital axis, by fitting the {\it
  Kepler} light curve we found $\io = 81.85 \pm 0.10$~degrees (Section
\ref{sec:transit}).  Knowledge of both the rotational and orbital
inclinations is not enough to determine the stellar
obliquity, because both measurements are subject to the usual
degeneracy $i \leftrightarrow 180^\circ - i$, and because we do not
know the position angle on the sky between the two axes.  Nevertheless
we may set a lower limit on the stellar obliquity of $|i_{\rm orb}-\is|=40\pm5$ (deg).
  
\begin{figure}[ht]
\centering
\includegraphics[width=0.8\columnwidth]{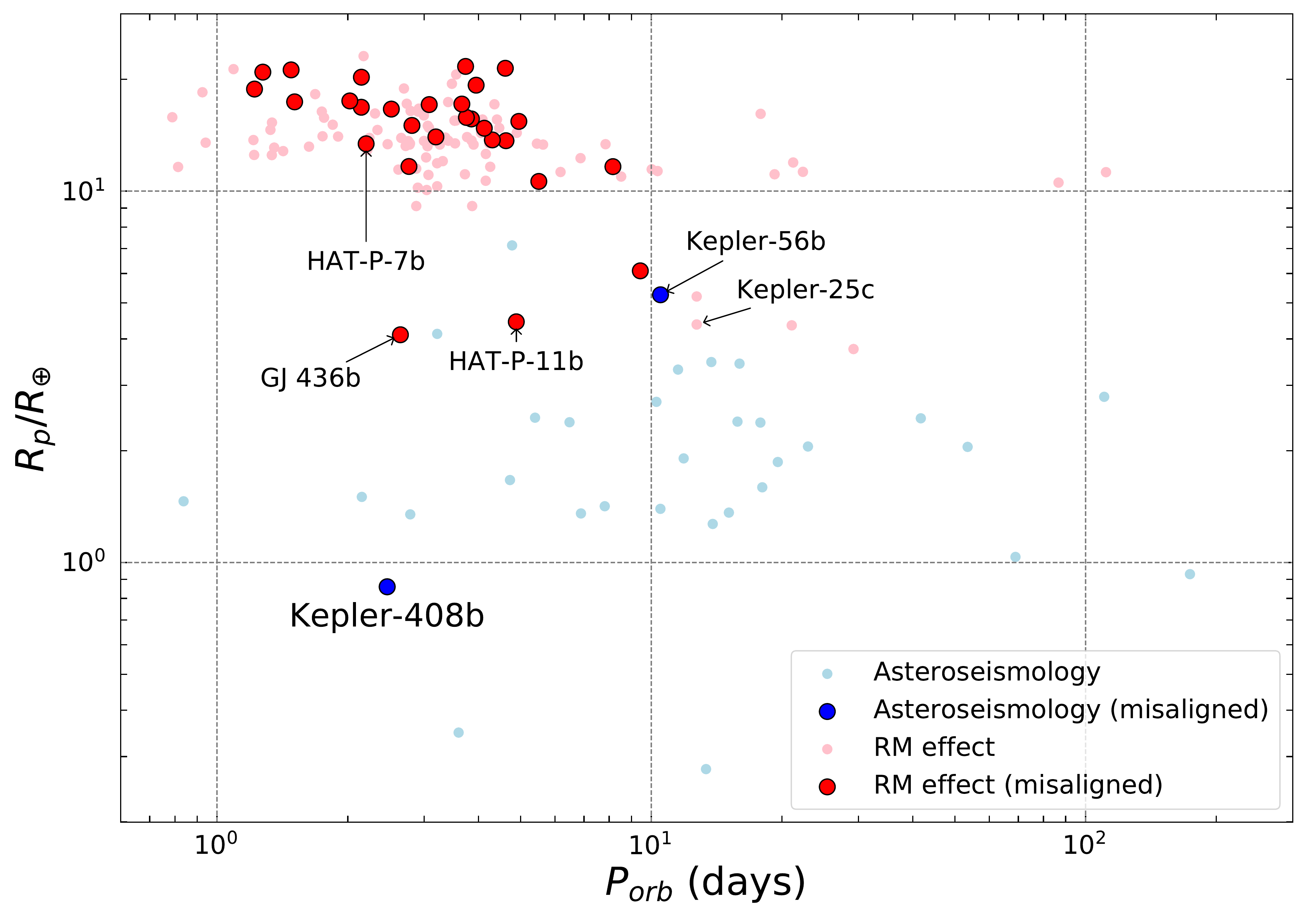}
\caption{ Sizes and orbital periods of planets for which the stellar
  obliquity has been constrained, based on the Rossiter-McLaughlin
  effect (red) and asteroseismology (blue).  Misaligned planets
    (with their $2\sigma$ lower limit of either $\lambda$ or $90^\circ
    -\is$ exceeding $30^{\circ}$) are marked in bold symbols.  Based
    on the compilation of \citet{Southworth2011} and our measurement
    \citep{Kamiaka+2018}.}
\label{fig:Porb-Rp}
\end{figure}

\begin{figure}[ht]
\centering
\includegraphics[width=0.8\columnwidth]{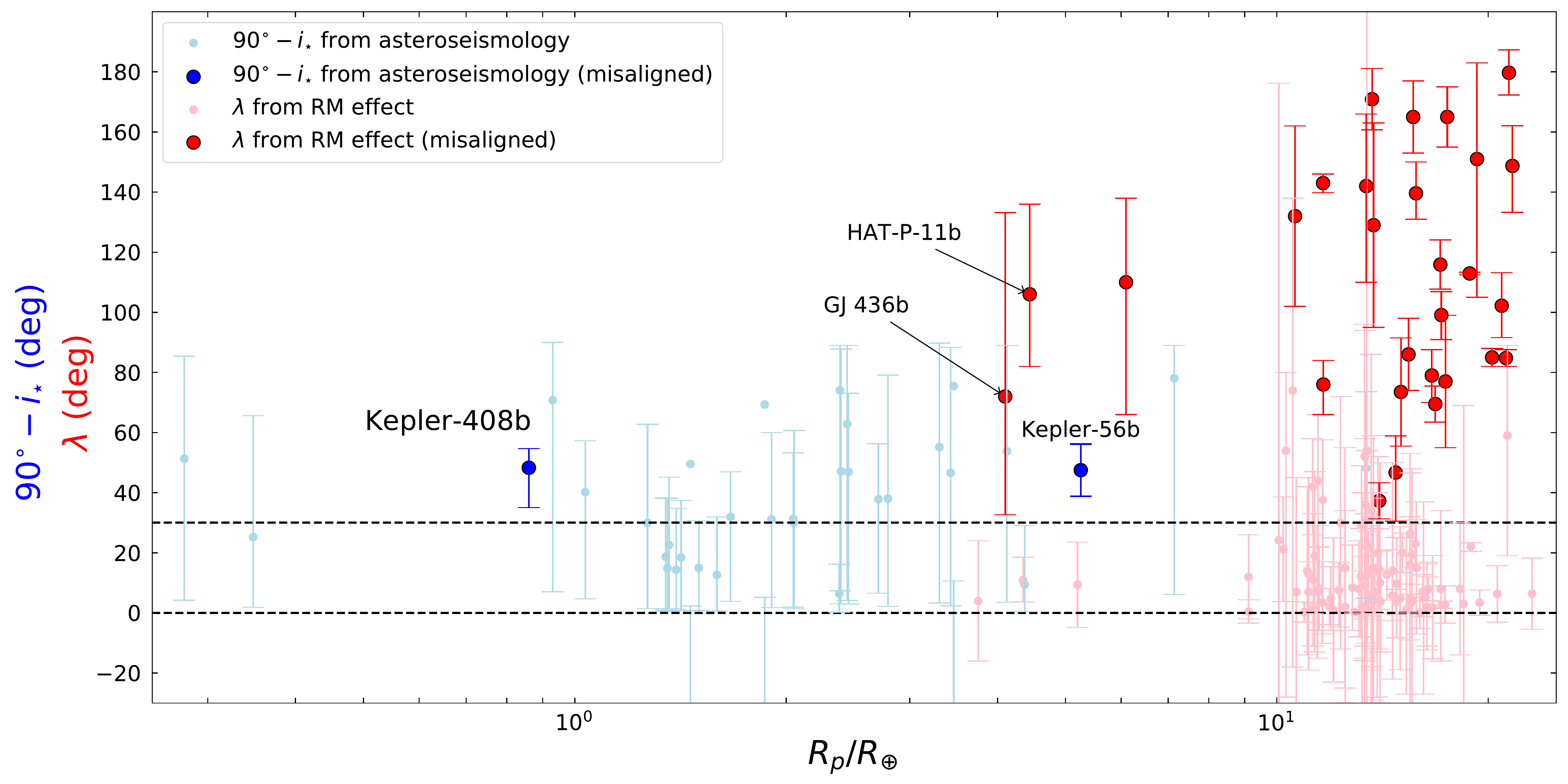}
\includegraphics[width=0.8\columnwidth]{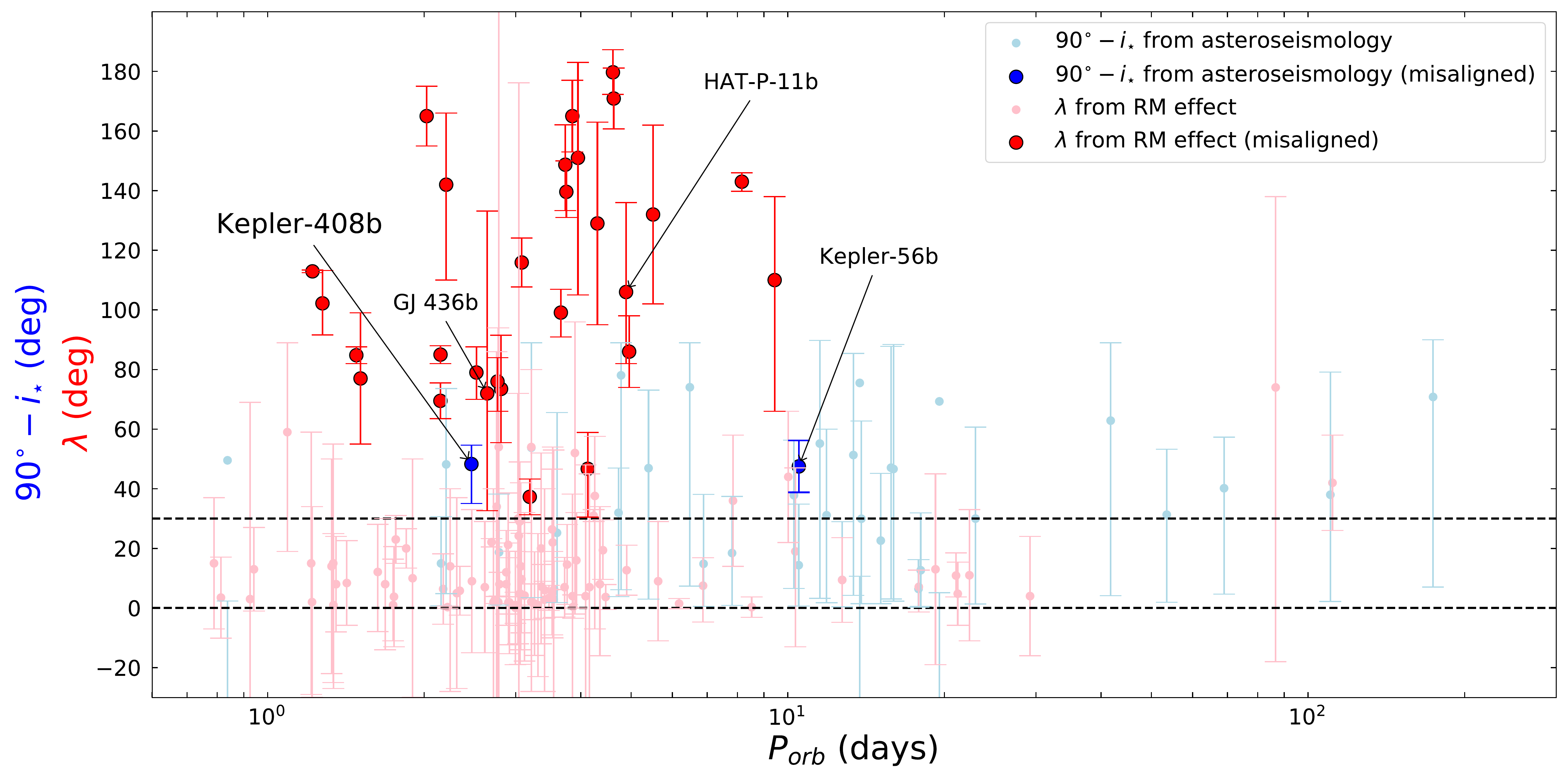}
\caption{ Observed spin-orbit angles for transiting exoplanets as a
  function of planetary radius and orbital period. Red symbols are for
  determinations of the position angle $\lambda$ based on the
  Rossiter-McLaughlin effect \citep{Southworth2011}. Blue symbols are
  for determinations of inclination based on asteroseismology
  \citep{Huber2013,Campante+2016,Kamiaka+2018}. For systems with more
  than one transiting planet, only the result for the innermost planet
  is plotted, with the sole exception of Kepler-25c.  The plotted
  error bars correspond to $2\sigma$ (95\%) confidence limits, and
  misaligned planets (with their $2\sigma$ lower limit of either
  $\lambda$ or $90^\circ -\is$ exceeding $30^{\circ}$) are marked in
  bold symbols. The horizontal dashed lines indicate 
    $30^\circ$ (our misalignment threshold) and $0^\circ$. Based
  on the compilation of \citet{Southworth2011} and our measurement
  \citep{Kamiaka+2018}.}
\label{fig:lambda_and_incl}
\end{figure}

Of all the planets known to have a spin-orbit misalignment,
Kepler-408b is the smallest by a factor of six, as illustrated in
Figures~\ref{fig:Porb-Rp} and \ref{fig:lambda_and_incl}.  As described
earlier, we classify systems as {\it misaligned} in those plots, if
either their sky-projected spin-orbit angle $\lambda$ or a proxy
  for the stellar obliquity in transiting systems, $90^\circ -\is$,
exceeds $30^{\circ}$ in 95\% confidence. The strong selection bias for
the RM measurement towards larger planets and shorter orbital periods
is clearly illustrated in the upper and lower panels of Figures
\ref{fig:lambda_and_incl}, in contrast to the homogeneous selection
for asteroseismic targets.

Those figures also identify other systems of particular
interest. Kepler-56 is an obliquely rotating star ($\is\sim 45^\circ$)
hosting two transiting planets \citep{Huber2013}.  HAT-P-7 and
Kepler-25 are the only known systems for which both
Rossiter-McLaughlin and asteroseismic measurements have been
successful \citep{Benomar+2014,Lund+2014}.  HAT-P-11b and GJ\,436b are
the smallest planets previously known to be misaligned
\citep{Winn+2010b,Yee2018,Bourrier+2018}.

Kepler-408 provides a clue about the origin of misalignments in general.
Stars and their planets are thought to form in a well-aligned state.
This is because the star and the protoplanetary disk inherit the same
direction of angular momentum from an initial clump of gas that
contracts under its own gravity.  The observation of a large obliquity
is an indication that something torqued the system out of alignment.
The circumstances and the timing of the torque are unknown.  Since all
the previous cases of large obliquities involved planets larger than
Neptune, some of the proposed theories have focused on giant planets.
The data have often been regarded as evidence that the formation of
close-orbiting giant planets, including hot Jupiters, involves
processes that tilt the planet's orbit \citep{Winn+2010,Triaud+2010}.

The case of Kepler-408 shows that orbit-tilting processes are not
specific to giant planets, and must occur at least occasionally for
``hot Earths.''  In a recently proposed theory for the formation of
very short-period terrestrial planets \citep{Petrovich+2018}, an inner
planet's orbital angular momentum is reduced through chaotic long-term
interactions with more distant planets, leading to spin-orbit
misalignments of $10^\circ$--$50^\circ$, as observed here. Another
theory involves a secular resonance with a more distant giant planet
\citep{HansenZink2015}, although in the case of Kepler-408, no
additional transiting planets are known.  The existing Doppler data do
not show any signals exceeding 4~m~s$^{-1}$ on timescales less than a
year \citep{Marcy2014}.  Other possibilities are that stars and their
protoplanetary disks are occasionally misaligned due to torques from
neighboring stars \citep{Batygin+2012} or that inner planets become
misaligned due to the torque from a wider-orbiting and misaligned
giant planet \citep{Lai+2018}. To decide among these and other
theories will require a larger and more diverse sample of planetary
systems for which the stellar obliquity can be probed.

\bigskip

\acknowledgments

We thank an anonymous referee for a careful reading of the earlier
manuscript and also for numerous valuable comments. We are grateful to
T.\ Campante for help in tracking down the reason for the discrepancy
with previously published results.  The numerical computation was
carried out on the DALMA cluster at New York University Abu Dhabi, and
the PC cluster at the Center for Computational Astrophysics, National
Astronomical Observatory of Japan. S.K.\ is supported by a Japan
Society for Promotion of Science (JSPS) Research Fellowship for Young
Scientists (No.\ 16J03121).  Y.S.\ gratefully acknowledges the support
from a Grant-in Aid for Scientific Research by JSPS No.\ 18H01247.
Work by F.D.\ and J.N.W.\ was supported by the Heising-Simons
Foundation.  O.B. thanks the invitation program supported by Research
Center for the Early Universe, the University of Tokyo.  The present
research was initiated during Y.S.'s visit at Princeton University
supported by the JSPS Core-to-Core Program ``International Network of
Planetary Sciences'', and also performed in part under contract with
the Jet Propulsion Laboratory (JPL) funded by NASA through the Sagan
Fellowship Program executed by the NASA Exoplanet Science Institute.

\bibliographystyle{aasjournal}

\end{document}